\documentclass[11pt]{article}
\usepackage[left=1in,top=1in,right=1in,bottom=1in,head=.1in,nofoot]{geometry}

\usepackage{setspace,url,bm,amsmath} %

\usepackage{titlesec} %
\titlelabel{\thetitle.\quad} %

\titleformat*{\section}{\bf\Large\center}

\usepackage{graphicx} %
\usepackage{bbm}
\usepackage{latexsym}
\usepackage{caption}
\usepackage[margin=20pt]{subcaption}
\usepackage{hyperref}
\usepackage{booktabs}
\usepackage{multirow}

\usepackage{enumitem}

\usepackage[table]{xcolor}

\newcommand{\GG}[1]{}

\usepackage{amsthm}
\usepackage{amssymb}
\usepackage{amsmath}
\usepackage{color}

\usepackage{comment}
\theoremstyle{definition}
\newtheorem{assumption}{Assumption}
\newtheorem*{theorem*}{Theorem}
\newtheorem{theorem}{Theorem}
\newtheorem*{rmk*}{Remark}

\newtheorem{example}{Example}

\newtheorem{condition}{Condition}

\newtheorem{remark}{Remark}

\newtheorem*{corollary*}{Corollary}

\def\treat{\text{1}}
\def\control{\text{0}}

\usepackage{natbib} %
\bibpunct{(}{)}{;}{a}{}{,} %

\usepackage{etoolbox} %
\apptocmd{\sloppy}{\hbadness 10000\relax}{}{} %

\usepackage{color}
\usepackage{listings}

\def\ind{\begin{picture}(9,8)
         \put(0,0){\line(1,0){9}}
         \put(3,0){\line(0,1){8}}
         \put(6,0){\line(0,1){8}}
         \end{picture}
        }
\def\nind{\begin{picture}(9,8)
         \put(0,0){\line(1,0){9}}
         \put(3,0){\line(0,1){8}}
         \put(6,0){\line(0,1){8}}
         \put(1,0){{\it /}}
         \end{picture}
    }

\def\Pr{\mathbb{P}}

\def\E{\mathbb{E}}

\def\ratio{R}

\makeatletter
\newcommand*{\bigcdot}{}%
\DeclareRobustCommand*{\bigcdot}{%
  \mathbin{\mathpalette\bigcdot@{}}%
}
\newcommand*{\bigcdot@scalefactor}{.5}
\newcommand*{\bigcdot@widthfactor}{1.15}
\newcommand*{\bigcdot@}[2]{%
  \sbox0{$#1\vcenter{}$}%
  \sbox2{$#1\cdot\m@th$}%
  \hbox to \bigcdot@widthfactor\wd2{%
    \hfil
    \raise\ht0\hbox{%
      \scalebox{\bigcdot@scalefactor}{%
        \lower\ht0\hbox{$#1\bullet\m@th$}%
      }%
    }%
    \hfil
  }%
}
\makeatother

\RequirePackage[normalem]{ulem}

\usepackage{soul}

\usepackage{subcaption}
\usepackage[textsize=tiny, textwidth = 2cm, shadow]{todonotes}

\begin{document}

\allowdisplaybreaks

\onehalfspacing

\title{\bf 
Sensitivity Analysis for Observational Studies with Flexible Matched Designs
}
\author{
Xinran Li
\footnote{
Xinran Li is Assistant Professor, Department of Statistics, University of Chicago, Chicago, IL 60637 (e-mail: \href{mailto:xinranli@uchicago.edu}{xinranli@uchicago.edu}). 
}
}
\date{}
\maketitle

\begin{abstract}
	Observational studies provide invaluable opportunities to draw causal inference, but they may suffer from biases due to pretreatment difference between treated and control units. Matching is a popular approach to reduce observed covariate imbalance. 
	To tackle unmeasured confounding, 
	a sensitivity analysis is often conducted to investigate how robust a causal conclusion is to the strength of unmeasured confounding. For matched observational studies, Rosenbaum proposed a sensitivity analysis framework that 
	uses the randomization of treatment assignments as the ``reasoned basis'' and imposes no model assumptions on the potential outcomes as well as their dependence on the observed and unobserved confounding factors. However, this otherwise appealing framework requires exact matching to guarantee its validity, which is hard to achieve in practice. 
	In this paper we provide an alternative inferential framework that shares the same procedure as Rosenbaum's approach but relies on a different justification.
	Our framework allows flexible matching algorithms
	and utilizes 
	alternative source of randomness, in particular random permutations of potential outcomes instead of treatment assignments, to guarantee statistical validity. 
\end{abstract}


\onehalfspacing

\section{Introduction}

Randomized experiments are the gold standard for drawing causal inference since they can produce comparable groups of units receiving different treatment arms. 
However, their implementation can be 
expensive, time-consuming and sometimes unethical. 
Observational studies provide invaluable opportunities for inferring causal effects, but at the same time they also bring new challenges due to confounding. 
Visible pretreatment differences between observed covariates in different treatment groups are called overt bias and can be handled by, e.g., matching, substratification, and weighting, and unobserved pretreatment differences are called hidden bias and 
need to be tackled in a different way
\citep{Rosenbaum1987}. 
A sensitivity analysis is a commonly used technique to tackle unmeasured confounding. 
It investigates how sensitive a causal conclusion is to the strength of unmeasured confounding \citep[][]{cornfield1959smoking}.

In the context of matched observational studies, \citet{Rosenbaum1987, Rosenbaum02a} proposed a sensitivity analysis framework considering biased randomization of treatment assignments within each matched set. 
For example, in a matched pair study, \citet{Rosenbaum1987} assumed that (i) units in different matched pairs are sampled independently, and (ii) within each matched pair, the treated unit is sampled from the population of treated units, while the control unit is independently sampled from the population of control units who have the same covariate value as the treated unit.
He then considered the conditional distribution of the treatment assignments 
among matched units, 
given their potential outcomes, observed covariates, unmeasured confounding and the matching structure. 
He showed that, in the resulting distribution, the treatment assignments are mutually independent across matched pairs, and moreover, in each matched pair, the probability that a unit receives the treatment with the other receiving control is proportional to the odds of its population-level treatment probability, which is the conditional probability of receiving treatment given its observed covariate and unmeasured confounding in the original study population. 
Finally, \citeauthor{Rosenbaum1987} imposed bounds on the strength of unmeasured confounding, quantified by the maximum odds ratio of the population-level treatment probabilities for units within each matched pair, and investigated how the inference, such as testing for 
the sharp null hypothesis 
of no treatment effect,
varies with the bounds. 

Rosenbaum's sensitivity analysis framework builds upon and extends the conventional randomization inference for randomized experiments, which uses random treatment assignments as the ``reasoned basis''
\citep{Fisher:1935}. 
However, Rosenbaum's framework requires exact matching, which is generally hard to achieve.  
Besides, the hypothesized data-generating process takes the matching structure as given, while in practice the matched data are usually constructed by applying some matching algorithm to a pool of units sampled from the study population. 
The resulting matching structure 
will depend on the  treatment assignments and covariates of the units in the pool, often in complicated ways. 
When matching is exact, 
Rosenbaum's randomization-based sensitivity analysis can still be justified when units are sampled independently prior to matching \citep[][Remark 1]{WL23}.
However,
when matching is inexact, 
Rosenbaum's analysis may be invalid \citep{Guo2022}.
This complication in post-matching inference has recently gained attention. 
\citet{Pashley2021} suggested that proper conditional inference for matched studies needs to consider the matching algorithm, 
and noted that different treatment assignments for matched units might yield a different matching structure, a phenomenon 
termed 
$Z$-dependence by \citet{pimentel2022covariate}.  
These 
highlight the need to
understand the conditional distribution of treatment assignments for matched units given the matching structure.
Such a task, however, can be 
challenging.

In this paper, we propose an alternative inferential framework for matched observational studies that explicitly 
takes into account the matching algorithm.  
We show that Rosenbaum's sensitivity analysis procedure can also be justified under this new framework, although the measure for bias will change and will depend explicitly on both the strength of unmeasured confounding and the quality of matching. 
This new framework focuses on the permutation distribution of potential outcomes given the treatment assignments, covariates, the matching algorithm, and within-set permutations of potential outcomes. 
The key advantage of this new framework is that, by conditioning on treatment assignments and covariates, we avoid the often intractable conditional distribution of treatment assignments given matching, which is required in Rosenbaum’s framework.
In sum, we can justify Rosenbaum's sensitivity analysis allowing flexible matching algorithms, but with modification on its interpretation. 
In addition, we can also incorporate matching quality into the sensitivity analysis, in a manner similar to \citet{pimentel2022covariate}.
It is also worth noting that the new framework does not justify all approaches within  Rosenbaum's framework; see \S \ref{sec:cov_adj}.

\section{Notation, Framework and Assumptions}

\subsection{Data prior to matching}
Consider an observational study with $N$ units and two treatment arms
prior to matching. 
For each unit $\ell \in [N]\equiv \{1,2,\ldots, N\}$, 
let $Y_\ell^{\treat}$ and $Y_\ell^{\control}$ 
denote its treatment and control potential outcomes, 
$X_\ell$ 
denote its pretreatment covariate vector, 
and 
$Z_\ell$ 
denote its treatment assignment, 
where $Z_\ell$ equals $1$ if the unit receives treatment and $0$ otherwise. 
The observed outcome is then $Y_\ell = Z_\ell Y_\ell^\treat + (1-Z_\ell) Y_\ell^\control$. 
We 
assume that the potential outcomes, covariates and treatment assignments are independent and identically distributed (i.i.d.) across all the $N$ units.
As discussed in Remarks A1 and A2 in the supplementary material, we can relax this assumption
by allowing the covariates to be fixed. 
Importantly, we cannot allow potential outcomes to be fixed constants as in finite population inference \citep{fcltxlpd2016}.  

\begin{assumption}\label{asmp:iid}
	$(Y_\ell^\treat, Y_\ell^\control, X_\ell, Z_\ell)$s for $ \ell\in [N]$ are i.i.d. following some distribution $\mathcal{P}$.   
\end{assumption}

Let $(Y^\treat, Y^\control, X, Z) \sim \mathcal{P}$.  
Let $p(\cdot \mid x)$ denote the density for the conditional distribution of $Y^\control$ given $X=x$, 
and 
$\pi_1(x, y) = 1 - \pi_0(x, y)$ denote the conditional probability of $Z=1$ given $(X, Y^\control) = (x,y)$.

\subsection{Matching}

We consider matching algorithms that match without replacement and produce 
one treated 
unit per matched set; 
see the supplementary material 
for extensions to matched studies with one treated or control unit per matched set. 
The matching algorithm, denoted by $\mathfrak{M}$, 
is allowed to 
depend on the observed treatment assignments and covariates 
in an arbitrary way. 
However, it is blind to observed outcomes \citep{Rubin2008trumps}. 
Let $I$ denote the number of matched sets, 
and $n_i$ denote the number of units within set $i\in [I]\equiv \{1,\ldots, I\}$. 
We index units within matched set $i$ by $i1, i2, \ldots, in_i$. 
Each matched unit $ij$ corresponds to a unit in the original data prior to matching, whose index is determined by the matching algorithm $\mathfrak{M}$. 

\subsection{Hypothesis testing for treatment effects and test statistics}\label{sec:test_stat}

We consider sharp null hypotheses such that, for each unit, as long as one of its potential outcomes is observed, we can impute the other. 
We focus on \citet{Fisher:1935}'s sharp null of no treatment effect, which can be written as 
$H_0: Y^1 - Y^0 = 0$ 
at the population level. 
See \S \ref{sec:test_inversion} for extensions to other sharp nulls.

We consider test statistics of form
$t(Z_{\bigcdot\bigcdot}, Y^0_{\bigcdot\bigcdot}) = \sum_{i=1}^I \sum_{j=1}^{n_i} Z_{ij} Q_{ij}, $
where $Z_{\bigcdot\bigcdot}$ and $Y^0_{\bigcdot\bigcdot}$ denote the vectors of treatment assignments and control potential outcomes for all matched units, and $Q_{ij}$s are functions of the control potential outcomes $Y^0_{\bigcdot\bigcdot}$. 
For example, 
\citet{Rosenbaummestimate2007} suggested the m-statistic with 
$Q_{ij} = n_i^{-1} \sum_{k\ne j} \psi( (Y_{ij}^0 - Y_{ik}^0)/s )$, 
where $\psi(\cdot)$ is an odd function and $s$ is,
say, the median of all the absolute pairwise differences between $Y_{ij}^0$s within each matched set. 
Recall that, under a sharp null hypothesis, the control potential outcomes $Y^0_{\bigcdot\bigcdot}$, as well as $Q_{ij}$s, can be imputed from the observed data.

\section{Matched pair studies}\label{sec:pair}

\subsection{Permutation inference versus randomization inference}

We consider first the matched pair studies, where each matched set is a pair containing one treated unit and one control unit. 
To conduct inference, 
we will consider the conditional null distribution of a test statistic 
given all the covariates, treatment assignments, the matching algorithm, 
and the within-pair permutations or equivalently order statistics of the control potential outcomes. 
That is, 
we condition on 
\begin{align}\label{eq:cond_set_perm}
\mathcal{F}_{\textup{pair}} 
& \equiv 
\{ (X_\ell, Z_\ell) \textup{ for } \ell\in [N], \ 
\mathfrak{M}, \ 
(Y_{i(1)}^0, Y_{i(2)}^0 )
\textup{ for } i \in [I] \},
\end{align}
where $Y_{i(1)}^0 \le Y_{i(2)}^0$ are sorted values of $Y_{i1}^0$ and $Y_{i2}^0$. 
In contrast, \citeauthor{Rosenbaum02a}
implicitly conditions on 
\begin{align}\label{eq:cond_set_rand}
\mathcal{R}_{\textup{pair}} 
& \equiv
\{X_\ell \textup{ for } \ell \in [N], \ 
\mathfrak{M} (\{(X_\ell, Z_\ell)\}_{\ell=1}^N), \ 
Y_{ij}^0 \textup{ for }  i \in [I] \text{ and } j=1,2, \ 
Z_{i1}+Z_{i2} \textup{ for }  i \in [I] \},
\end{align}
in addition to other unmeasured confounders. 
For convenience, we will refer to our framework as permutation-based and Rosenbaum's as randomization-based, despite that \citet{Rosenbaum1987} also used the term ``permutation inference''. 
Accordingly, we will refer to the conditional distribution of a certain quantity given $\mathcal{F}_{\textup{pair}}$ as its \textit{permutation distribution}, and that given $\mathcal{R}_{\textup{pair}}$ as its \textit{randomization distribution}.

The key difference between our and Rosenbaum's frameworks comes from the difference in the conditioning sets in \eqref{eq:cond_set_perm} and \eqref{eq:cond_set_rand}. 
Consider the test statistic in \S \ref{sec:test_stat}. 
Its randomness 
conditional on $\mathcal{F}_{\textup{pair}}$ comes solely from the random permutation of potential outcomes within each pair, while its randomness conditional on $\mathcal{R}_{\textup{pair}}$ comes solely from the randomization of treatment assignments for matched units. 
Due to potentially complicated dependence of the matching algorithm $\mathfrak{M}$ on the treatment assignments in the original data, it can be quite challenging to tackle the conditional distribution of $Z_{ij}$s for matched units given $\mathcal{R}_{\textup{pair}}$. 
As detailed below, the key advantage of our framework is that the conditional distribution of the test statistic given  $\mathcal{F}_{\textup{pair}}$ can be characterized more easily, regardless of the complexity of matching.

\subsection{Matching quality, unmeasured confounding strength, and 
biases from uniform 
permutation}\label{sec:bias_pair}

Once conditional on $\mathcal{F}_{\textup{pair}}$, the randomness in a test statistic  
in \S \ref{sec:test_stat} comes solely from the random potential outcomes $Y_{ij}^0$s and more precisely their within-pair random permutations. 
For clarity and without loss of generality, we assume that $Z_{i1} = 1$ and $Z_{i2} = 0$ for all matched pair $i\in [I]$. 
From Assumption \ref{asmp:iid}, 
we can verify that, conditional on $\mathcal{F}_{\textup{pair}}$, $Y_{i\bigcdot}^0$s are mutually independent across all matched pairs. 
Furthermore, for each pair $i$, conditional on $\mathcal{F}_{\textup{pair}}$, 
the potential outcomes $Y_{i\bigcdot}^0 = (Y_{i1}^0, Y_{i2}^0)$ 
are either $(Y_{i(2)}^0, Y_{i(1)}^0)$ or $(Y_{i(1)}^0, Y_{i(2)}^0)$. 
The ratio of the probabilities of taking these two values has the following equivalent forms:
\begin{equation}\label{eq:decomp_ps}
\ratio_i
= 
\frac{
	\Pr( Y_{i1}^0 = Y_{i(2)}^0 \mid \mathcal{F}_{\textup{pair}})
}{
	\Pr( Y_{i1}^0 = Y_{i(1)}^0 \mid \mathcal{F}_{\textup{pair}})
}
= 
\underbrace{
	\frac{
		p(Y_{i(2)}^0 \mid X_{i1}) / p(Y_{i(2)}^0 \mid X_{i2})
	}{
		p(Y_{i(1)}^0 \mid X_{i1}) / p(Y_{i(1)}^0 \mid X_{i2})
	}
}_\textup{matching quality $\ratio_{\text{m}i}$}\underbrace{\frac{
		\pi_0(X_{i2}, Y_{i(1)}^0)/\pi_0(X_{i2}, Y_{i(2)}^0)
	}{
		\pi_1(X_{i1}, Y_{i(1)}^0)/\pi_1(X_{i1}, Y_{i(2)}^0)
	}
}_\textup{unmeasured confounding strength $\ratio_{\text{u}i}$}, 
\end{equation}
see the supplementary material for the derivation. 
The 
expression in
\eqref{eq:decomp_ps} is decomposed into two terms: $\ratio_{\text{m}i}$ for the matching quality and $\ratio_{\text{u}i}$ for the unmeasured confounding strength. 
Moreover, 
$\ratio_{\text{m}i}$ 
reduces to $1$ if matching is exact, 
and $\ratio_{\text{u}i}$ 
reduces to $1$ if there is no unmeasured confounding.

We then define 
$\Gamma_i = \max\{ \ratio_{i}, \ratio_{i}^{-1} \}$
for all $i$. 
Obviously, $\Gamma_i \ge 1$, and it becomes $1$ if matching is exact and there is no unmeasured confounding. 
If $\Gamma_i = 1$, then, conditional on $\mathcal{F}_{\textup{pair}}$, 
$Y_{i\bigcdot}^0$ is equally likely to be 
the two permutations 
of $(Y_{i(1)}^0, Y_{i(2)}^0)$;
otherwise, one permutation is $\Gamma_i$-times more likely than the other.  
Thus, $\Gamma_i$ quantifies the bias from uniform permutation,
which is achieved at the ideal scenario with exact matching and no unmeasured confounding. 
Analogously, 
when studying the randomization distribution of treatment assignments, 
\citet{Rosenbaum02a} introduced a similar $\Gamma$ parameter to characterize the bias from uniform randomization, which is also achieved at the same ideal scenario. 
Thus, 
our $\Gamma_i$s play a similar role as Rosenbaum's $\Gamma$, measuring the deviations from the ideal scenario.  
However, their expressions are different. 
Specifically, 
the $\Gamma_i$s 
from \eqref{eq:decomp_ps}
contain two distinct components: unmeasured confounding strength and matching quality. The latter 
is absent 
in Rosenbaum's $\Gamma$ because he focused on exact matching.

\subsection{The exact permutation distributions of test statistics}

We now study the exact permutation distribution of a test statistic. 
We impose the following 
``symmetry'' 
condition,  
which holds for many commonly used test statistics, such as the m-statistics discussed in \S \ref{sec:test_stat}.

\begin{condition}\label{cond:symmetry}
The test statistic $t(Z_{\bigcdot\bigcdot}, Y^0_{\bigcdot\bigcdot}) = \sum_{i=1}^I \sum_{j=1}^{2} Z_{ij} Q_{ij}$ satisfies that,  
for all $ij$, 
$Q_{ij}$, as a function of the potential outcomes $Y^0_{\bigcdot\bigcdot}$, has the following form for some constant function $\phi_i$: 
$
Q_{ij} = \phi_{i}( 
Y_{ij}^0,
(Y_{1(1)}^0, Y_{1(2)}^0), 
\ldots, 
(Y_{I(1)}^0, Y_{I(2)}^0)
), 
$
where the second until the last arguments depend only on 
permutations or equivalently order statistics of the two control potential outcomes within each matched pair. 
\end{condition}

For all $ij$, 
let 
$
q_{ij} = \phi_{i}( 
Y_{i{(j)}}^0, 
(Y_{1(1)}^0, Y_{1(2)}^0), 
\ldots, 
(Y_{i(1)}^0, Y_{i(2)}^0)
),
$
with $\phi_i$ defined as in 
Condition \ref{cond:symmetry}.

\begin{theorem}\label{thm:exact_pair}
Under Assumption \ref{asmp:iid} and $H_0$, 
the permutation distribution of any test statistic $t(Z_{\bigcdot\bigcdot}, Y^0_{\bigcdot\bigcdot})$ 
satisfying Condition \ref{cond:symmetry}
is  
$
t(Z_{\bigcdot\bigcdot}, Y^0_{\bigcdot\bigcdot}) 
\mid \mathcal{F}_{\textup{pair}}
\sim 
\sum_{i=1}^I \widetilde{Q}_i, 
$
where $\widetilde{Q}_i$s are mutually independent, 
and for each $i$, 
$\widetilde{Q}_i$ equals $q_{i1}$ with probability $1/(1+\ratio_i)$ and $q_{i2}$ with probability $\ratio_i/(1+\ratio_i)$, with $\ratio_i$ defined in \eqref{eq:decomp_ps}. 
\end{theorem}

As discussed shortly, 
we will impose bounds on the $\ratio_i$s or $\Gamma_i$s, 
which 
constrain 
the matching quality and unmeasured confounding strength. 
We then derive the corresponding worst-case permutation distribution of the test statistic, 
yielding conservative yet valid $p$-values for $H_0$. 
We finally investigate how the inference varies as our bounds change.  
This is in the same spirit as Rosenbaum's sensitivity analysis. 

\subsection{Connecting permutation and randomization distributions}\label{sec:link_pair}

To conduct sensitivity analysis with bounds on the biases $\Gamma_i$s, 
we will first connect the permutation distribution in Theorem \ref{thm:exact_pair} to a randomization distribution, 
based on which we can then utilize Rosenbaum's approach to derive worst-case permutation distributions as well as valid $p$-values for testing $H_0$.
Define a random vector $A_{\bigcdot\bigcdot} \equiv (A_{11}, \ldots, A_{I2})$ such that 
$(A_{i1}, A_{i2})$s are mutually independent across all $i$,
and for each $i$, 
$A_{i1} + A_{i2}=1$  
and 
$\Pr(A_{i1} = 1) = 1/(1+\ratio_i)$.  
We can show 
there exist $u_{ij}$s in $[0,1]$
such that 
\begin{align}\label{eq:dist_A}
\Pr(A_{\bigcdot\bigcdot} = a_{\bigcdot\bigcdot})
& = 
\prod_{i=1}^I \frac{\exp(\gamma_i \sum_{j=1}^{2} a_{ij} u_{ij})}{\sum_{j=1}^{2} \exp(\gamma_i u_{ij})}
\quad \textup{ with } 
\gamma_i = \log \Gamma_i \ge 0 \textup{ for all } i
\end{align}
if $a_{ij}$s are all in $\{0,1\}$ and $a_{i1} + a_{i2} = 1$ for all $i$,
and $\Pr(A_{\bigcdot\bigcdot} = a_{\bigcdot\bigcdot}) = 0$ otherwise. 
We can view the $A_{ij}$s as random treatment assignments from a paired randomized experiment.

\begin{theorem}\label{thm:link_perm_rand}
The permutation distribution of the test statistic in Theorem \ref{thm:exact_pair} 
is the same as the randomization distribution of $t(A_{\bigcdot\bigcdot}, Y^0_{\bigcdot(\bigcdot)}) = 
\sum_{i=1}^I \sum_{j=1}^2 A_{ij} q_{ij},
$
where $Y^0_{\bigcdot(\bigcdot)} = (Y_{1(1)}^0, 
\ldots, Y_{I(2)}^0)$ are treated as fixed, $q_{ij}$s are defined as in Theorem \ref{thm:exact_pair},   
and $A_{\bigcdot\bigcdot}$ follows the distribution in \eqref{eq:dist_A} for some $u_{ij}$s in $[0,1]$. 
\end{theorem}

Below we give some intuition for the equivalence in Theorem \ref{thm:link_perm_rand}. In our permutation-based framework, we consider the likelihood of different permutations of the potential outcomes given fixed treatment assignments. In contrast, in the randomization-based framework, we consider the likelihood of different permutations of the treatment assignments given fixed potential outcomes.
These two frameworks are similar in the sense that they both consider the likelihood of different pairings between potential outcomes and treatment assignments. 
In addition, the equivalent form in \eqref{eq:decomp_ps} can help us understand the likelihood of different permutations of potential outcomes through the perspective of treatment assignment mechanism.

Theorem \ref{thm:link_perm_rand} has important implications. 
It shows that the  permutation distribution of a test statistic can be equivalently viewed as a randomization distribution of the same test statistic under a hypothetical paired randomized experiment. 
Moreover, 
the treatment assignment mechanism of 
this hypothetical experiment
has the same form as that under Rosenbaum's sensitivity analysis; see, e.g., \citet[][equation (4.7)]{Rosenbaum02a}. 
Consequently, following \citet{Rosenbaum1987} and \citet{WL23}, we can conduct valid tests for $H_0$ with a uniform bound on the biases $\Gamma_i$s or
with 
bounds on quantiles of the biases $\Gamma_i$s. 
We 
can then investigate how the inferred causal conclusions
change as the 
bounds on 
the biases $\Gamma_i$s
vary. 

In sum, the sensitivity analysis under our permutation-based framework shares the same procedure as that under Rosenbaum's randomization-based framework. 
However, their interpretation and justification, as well as their bias measures, are considerably different. 
For conciseness, 
we omit detailed implementation 
of such sensitivity analyses and refer readers to \citet{Rosenbaum1987, Rosenbaum02a} and \citet{WL23}.

\subsection{Adaptive permutation inference and its connection to adaptive randomization inference}\label{sec:adapt_pair}

The bias measure in \eqref{eq:decomp_ps}
depends on both matching quality and unmeasured confounding strength. 
Below we discuss a strategy to estimate the matching quality $\ratio_{\text{m}i}$, so that the analysis can explicitly take into account covariate imbalance after matching and focus more on the strength of unmeasured confounding. 
Suppose we also have some external samples, which are independent of the study data used for matching. 
Let $\hat{p}(\cdot \mid x)$ denote an estimate for the conditional density of $Y^0$ given $X=x$, constructed from external samples, where  the control potential outcomes for treated units are imputed based on their observed outcomes and the sharp null hypothesis under investigation.
For each $i\in [I]$, we then estimate the matching quality $\ratio_{\text{m}i}$ in  \eqref{eq:decomp_ps} by $\hat{\ratio}_{\text{m}i}$, which has the same expression as $\ratio_{\text{m}i}$ but with $p(\cdot \mid \cdot)$ replaced by $\hat{p}(\cdot \mid \cdot)$. 
Define $\tilde{\ratio}_{\textup{u}i} = \ratio_i/ \hat{\ratio}_{\text{m}i} = \ratio_{\text{u}i} \ratio_{\text{m}i}/\hat{\ratio}_{\text{m}i}$ 
and $\tilde{\Gamma}_{\textup{u}i} = \max\{\tilde{\ratio}_{\textup{u}i}, \tilde{\ratio}_{\textup{u}i}^{-1}\}$. 
When our density estimator is close to the truth, for each $i$, 
we have $\hat{\ratio}_{\text{m}i} \approx \ratio_{\text{m}i}$ and consequently $\tilde{\ratio}_{\textup{u}i} \approx \ratio_{\text{u}i}$. 
Hence, 
compared to the biases $\Gamma_i$s in \S \ref{sec:bias_pair}, 
the biases $\tilde{\Gamma}_{\textup{u}i}$s here focus more on the strength of unmeasured confounding. 

Below we consider bounding the biases $\tilde{\Gamma}_{\text{u}i}$s, which mainly constrain the strength of unmeasured confounding, 
and investigate how sensitive the inference results will be.  
This is distinct from the sensitivity analysis in \S \ref{sec:link_pair}, where the imposed bounds are on the $\Gamma_i$s and constrain both unmeasured confounding strength and matching quality. 
Moreover, $\Gamma_i$s measure biases from uniform permutations, whereas $\tilde{\Gamma}_{\textup{u}i}$s measure biases from nonuniform permutations, under which $Y_{i\bigcdot}^0$ equals $(Y_{i(1)}^0, Y_{i(2)}^0)$ with probability $1/(1+\hat{\ratio}_{\text{m}i})$ and $(Y_{i(2)}^0, Y_{i(1)}^0)$ with probability $\hat{\ratio}_{\text{m}i}/(1+\hat{\ratio}_{\text{m}i})$, independently across all $i\in [I]$.

We first connect 
the permutation distribution in Theorem \ref{thm:exact_pair} to an adaptive randomization distribution \citep{pimentel2022covariate}.  
For $u_{ij}$s in $[0,1]$, 
define a random vector $A_{\bigcdot\bigcdot} \equiv (A_{11}, \ldots, A_{I2})$ such that 
\begin{align}\label{eq:cov_adap_pair}
\Pr(A_{\bigcdot\bigcdot} = a_{\bigcdot\bigcdot})
& = 
\prod_{i=1}^I \frac{\exp\{ a_{i1} \tilde{\gamma}_{\text{u}i} u_{i1} + a_{i2} (\log \hat{\ratio}_{\text{m}i}+\tilde{\gamma}_{\text{u}i} u_{i2})\}}{
	\exp(\tilde{\gamma}_{\text{u}i} u_{i1}) + \exp(\log \hat{\ratio}_{\text{m}i}+ \tilde{\gamma}_{\text{u}i} u_{i2})
}
\quad \textup{ with } 
\tilde{\gamma}_{\text{u}i} = \log \tilde{\Gamma}_{\text{u}i} \ge 0 \textup{ for all } i
\end{align}
if $a_{ij}$s are all in $\{0,1\}$ and $a_{i1} + a_{i2} = 1$ for all $i$,
and $\Pr(A_{\bigcdot\bigcdot} = a_{\bigcdot\bigcdot}) = 0$ otherwise. 

\begin{theorem}\label{thm:link_perm_rand_adaptive}
The permutation distribution of the test statistic in Theorem \ref{thm:exact_pair} 
is the same as the randomization distribution of $t(A_{\bigcdot\bigcdot}, Y^0_{\bigcdot(\bigcdot)}) = 
\sum_{i=1}^I \sum_{j=1}^2 A_{ij} q_{ij},
$
where $Y^0_{\bigcdot(\bigcdot)} = (Y_{1(1)}^0, 
\ldots, Y_{I(2)}^0)$ is treated as fixed, $q_{ij}$s are defined as in Theorem \ref{thm:exact_pair},   
and $A_{\bigcdot\bigcdot}$ follows the distribution in \eqref{eq:cov_adap_pair} for some $u_{ij}$s in $[0,1]$. 
\end{theorem}

Importantly, Theorem \ref{thm:link_perm_rand_adaptive} allows us to derive the worst-case permutation distribution of a test statistic under a uniform bound on the $\tilde{\Gamma}_{\text{u}i}s$ in a similar way as \citet{pimentel2022covariate}; see the supplementary material for details.
Despite the connection in Theorem \ref{thm:link_perm_rand_adaptive}, our approach differs significantly from that of \citet{pimentel2022covariate}.
First, the baseline permutation distribution when $\tilde{\Gamma}_{\text{u}i}$s are $1$ for all $i$ is obtained using estimated density of $Y^0$ given $X$, while the baseline randomization distribution in \citet{pimentel2022covariate} is obtained using estimated propensity score. Both baseline distributions rely crucially on the covariate imbalance within each matched pair. 
Second, when the unmeasured confounding strength is bounded by some $\Gamma>1$, ignoring the estimation uncertainty, 
the biases from the baseline permutation distribution under our framework are at most $\Gamma$, whereas the biases from the baseline randomization distribution in \citet{pimentel2022covariate} are bounded instead by $\Gamma^2$. 
In addition to difference in the definition of unmeasured confounding strength, one reason is that our baseline permutation distribution depends only on the (estimated) conditional distribution of $Y^0$ given $X$, which is variationally independent of the true treatment probability $\pi_1(x,y)$.
This, however, is no longer true in \citet{pimentel2022covariate}'s framework, since the baseline randomization distribution depends on the (estimated) propensity score $\Pr(Z=1\mid X) = \E \{ \pi_1(X,Y^0) \mid X \}$, which is related to the true treatment probability.  
Finally, our permutation-based adaptive sensitivity analysis allows flexible matching algorithms and does not suffer from the Z-dependency issue, which, however, may affect the analyses in \citet{pimentel2022covariate}.

\section{Matched set studies}\label{sec:set}

\subsection{Matching quality, unmeasured confounding strength, and biases from uniform permutation}\label{sec:bias_set}

We now consider more general matched studies, where each matched set contains one treated unit and one or multiple control units; see the supplementary material for extensions to matched studies with one treated or one control unit per set. 
Similar to \S \ref{sec:pair}, 
we consider the conditional distribution of a test statistic 
given 
\begin{align*}
\mathcal{F}_{\textup{set}} 
\equiv 
\{ (X_\ell, Z_\ell) \textup{ for } \ell \in N, \ \mathfrak{M}, \ 
Y_{i (\bigcdot)}^0 \textup{ for } i \in [I] \},
\end{align*}
where $Y_{i (\bigcdot)}^0 = (Y_{i (1)}^0, Y_{i (2)}^0, \ldots, Y_{i (n_i)}^0)$ consists of the sorted values of $Y_{ij}^0$s for matched set $i$.
Without loss of generality, we also 
assume that $Z_{i1} = 1$ and $Z_{i2} = \ldots = Z_{i n_i} = 0$ for all matched set $i$. 
We refer to the conditional distribution of a 
quantity
given $\mathcal{F}_{\textup{set}}$ again as a permutation distribution.

The permutation distribution of a test statistic 
depends crucially on the randomness of the potential outcomes.  
Under Assumption \ref{asmp:iid} and conditional on $\mathcal{F}_{\textup{set}}$, the potential outcomes are mutually independent across all matched sets, and, for each matched set $i$, the potential outcomes $Y_{i\bigcdot}^0 = (Y_{i1}^0, \ldots, Y_{in_i}^0)$ take values in $\{ Y_{i(\sigma)}^0 = (Y_{i(\sigma_1)}^0, Y_{i(\sigma_2)}^0, \ldots, Y_{i(\sigma_{n_i})}^0): \sigma \in \mathcal{S}_{n_i} \}$, where $\mathcal{S}_{n_i}$ denotes the set of all permutations of $(1, 2, \ldots, n_i)$.
As demonstrated in the supplementary material, 
we can show that 
\begin{align*}
\Pr(Y_{i\bigcdot}^0 = Y_{i(\sigma)}^0 \mid \mathcal{F}_{\text{set}}) 
& 
\propto
\prod_{k=1}^{n_i} p(Y_{i(\sigma_k)}^0 \mid X_{ik})
\cdot \pi_1(X_{i1}, Y_{i(\sigma_1)}^0) \prod_{k=2}^{n_i} \pi_0(X_{ik}, Y_{i(\sigma_k)}^0)
\equiv p_{i\sigma} \cdot \pi_{i\sigma}, 
\end{align*}
where $p_{i\sigma}$ denotes the product of the $n_i$ terms related to the conditional distribution of $Y^0$ given $X$, and $\pi_{i\sigma}$ denotes the product of the $n_i$ terms related to the conditional distribution of $Z$ given $(X, Y^0)$. 
Below we focus on the permutation distribution of  $Y_{i1}^0$s, which is sufficient for 
characterizing the permutation distribution of a test statistic under our consideration. 
The exact permutation distribution of $Y_{i1}^0$ may have rather complicated forms, involving sums over permutations in $\mathcal{S}_{n_i}$. 
We instead bound the ratio of the probabilities of $Y_{i1}^0$ taking any two possible values, 
based on which we can have a set of distributions covering the true permutation distribution of $Y_{i1}^0$.  
Let  $p_{\text{m}ij} = \sum_{\sigma \in \mathcal{S}_{n_i}: \sigma_1 = j} p_{i\sigma}$ for $j\in [n_i] \equiv \{1,\ldots, n_i\}$, 
\begin{align}\label{eq:match_quality_set}
\Gamma_{\text{m}i} \equiv
\max_{j,k\in [n_i]}\frac{ p_{\text{m}ij}  }{
	p_{\text{m}ik}
}
\le \max_{j,k, l\in [n_i]} 
\ 
\frac{
	p(Y_{i(j)}^0 \mid X_{i1}) / p(Y_{i(j)}^0 \mid X_{il})
}{
	p(Y_{i(k)}^0 \mid X_{i1}) / p(Y_{i(k)}^0 \mid X_{il})
},
\quad (i\in [I])
\end{align}
which measures the matching quality and reduces to $1$ in the case of exact matching, and 
\begin{align}\label{eq:confound_strength_set}
\Gamma_{\text{u}i} \equiv \frac{
	\max_{\sigma \in \mathcal{S}_{n_i}} \pi_{i\sigma} 
}{
	\min_{\psi \in \mathcal{S}_{n_i}} \pi_{i\psi} 
} \le 
\max_{j,k\in [n_i]}\frac{ \pi_1(X_{i1}, Y_{i(j)}^0) }{\pi_1(X_{i1}, Y_{i(k)}^0)} \cdot 
\prod_{l=2}^{n_i} \left[ \max_{j,k\in [n_i]}\frac{ \pi_0(X_{il}, Y_{i(j)}^0) }{\pi_0(X_{il}, Y_{i(k)}^0)} \right],
\quad (i \in [I])
\end{align}
which measures the unmeasured confounding strength and reduces to $1$ in the absence of unmeasured confounding. 
Define further $\Gamma_i \equiv \Gamma_{\text{m}i} \Gamma_{\text{u}i}$ for all $i \in [I]$. 
As demonstrated in the supplementary material, 
\begin{align}\label{eq:set_prob_ratio}
\frac{\Pr(Y_{i1}^0 = Y_{i(j)}^0\mid \mathcal{F}_{\textup{set}})}{\Pr(Y_{i1}^0 = Y_{i(k)}^0 \mid \mathcal{F}_{\textup{set}})}
& 
\in 
\frac{ p_{\text{m}ij}  }{
	p_{\text{m}ik} 
}
\cdot 
\big[ 
\Gamma_{\text{u}i}^{-1}, \  
\Gamma_{\text{u}i}
\big]
\subset 
\big[ 
\Gamma_i^{-1}, \  
\Gamma_i
\big], 
\quad 
\text{ for all } i \in [I] \text{ and } j,k\in [n_i]. 
\end{align}
In the supplementary material, we provide tighter bounds for the probability ratio in \eqref{eq:set_prob_ratio}, aligning more closely with \citet{Rosenbaum02a}’s measure of unmeasured confounding strength via treatment probability odds ratios. 
We use the bounds in \eqref{eq:set_prob_ratio}
for their convenience in the later adaptive permutation inference.

The bias $\Gamma_i$ in \eqref{eq:set_prob_ratio} is determined by the matching quality in \eqref{eq:match_quality_set} and unmeasured confounding strength in \eqref{eq:confound_strength_set}. 
In the ideal scenario with exact matching and no measured confounding, $\Gamma_i=1$, and the permutation distribution of $Y_{i1}^0$ is uniform on $\{ Y_{i(1)}^0, \ldots, Y_{i(n_i)}^0 \}$. 
Thus, intuitively, $\Gamma_i$ quantifies the bias from uniform permutation, which is achieved at the ideal scenario. 
In the supplementary material, we further simplify the biases
in the case of exact matching or no unmeasured confounding.

\subsection{The permutation distributions of test statistics and their connection to randomization distributions}\label{sec:sen_base_unif_perm}

We now study the permutation distribution of a test statistic. Similar to Condition \ref{cond:symmetry}, we impose some ``symmetry'' condition on the test statistic.
The m-statistics in \S \ref{sec:test_stat}, for example, satisfy this condition. 

\begin{condition}\label{cond:symmetry_set}
The statistic $t(Z_{\bigcdot \bigcdot}, Y_{\bigcdot \bigcdot}^0) = \sum_{i=1}^I \sum_{j=1}^{n_i} Z_{ij} Q_{ij}$ satisfies that, for all $ij$, 
$Q_{ij}$, as a function of the potential outcomes $Y_{\bigcdot \bigcdot}^0$, has the following form for some constant function $\phi_i$: 
$
Q_{ij} = \phi_i( Y_{i(j)}^0, 
Y_{1(\bigcdot)}^0, 
\ldots, 
Y_{I(\bigcdot)}^0
), 
$
where the second to last arguments 
depend only on the sorted values of $Y_{i\bigcdot}^0$s. 
\end{condition}

Let
$
q_{ij} = \phi_i( Y_{i(j)}^0, Y_{1(\bigcdot)}^0, 
\ldots, 
Y_{I(\bigcdot)}^0
)
$
for all $ij$, with $\phi_i$ defined as in Condition \ref{cond:symmetry_set}.

\begin{theorem}\label{thm:perm_set}
Under Assumption \ref{asmp:iid} and $H_0$, for any test statistic $t(Z_{\bigcdot\bigcdot}, Y_{\bigcdot\bigcdot}^0)$ satisfying Condition \ref{cond:symmetry_set}, 
its permutation distribution is
$
t(Z_{\bigcdot \bigcdot}, Y_{\bigcdot \bigcdot}^0) \mid \mathcal{F}_{\textup{set}}
\ \sim \ 
\sum_{i=1}^I \widetilde{Q}_i, 
$
where $\widetilde{Q}_i$s are mutually independent,
and, for each $i$, $\widetilde{Q}_i$ takes value in $\{q_{i1}, \ldots, q_{in_i}\}$ and $\Gamma_i^{-1} \le \Pr(\widetilde{Q}_i = q_{ij})/\Pr(\widetilde{Q}_i = q_{ik}) \le \Gamma_i$ for all $j,k \in [n_i]$. 
\end{theorem}

Below we connect the permutation distribution of a test statistic to its randomization distribution under a hypothetical experiment. 
For any $u_{ij}$s in $[0,1]$, 
define a random vector $A_{\bigcdot\bigcdot} \equiv (A_{11}, \ldots, A_{In_I})$ such that 
\begin{align}\label{eq:dist_A_set}
\Pr(A_{\bigcdot\bigcdot} = a_{\bigcdot\bigcdot})
& = 
\prod_{i=1}^I \frac{\exp(\gamma_i \sum_{j=1}^{n_i} a_{ij} u_{ij})}{\sum_{j=1}^{n_i} \exp(\gamma_i u_{ij})}
\quad \text{with } \gamma_i = \log \Gamma_i \ge 0 \text{ for all } i
\end{align}
if $a_{ij}$s are all in $\{0,1\}$ and $\sum_{j=1}^{n_i} a_{ij} = 1$ for all $i$, and $\Pr(A_{\bigcdot\bigcdot} = a_{\bigcdot\bigcdot}) = 0$ otherwise. 

\begin{theorem}\label{thm:link_perm_rand_set}
The permutation distribution of the test statistic in Theorem \ref{thm:perm_set} is the same as the randomization distribution of 
$
t(A_{\bigcdot\bigcdot}, Y^0_{\bigcdot(\bigcdot)}) = 
\sum_{i=1}^I \sum_{j=1}^{n_i} A_{ij} q_{ij},
$
where $Y^0_{\bigcdot(\bigcdot)} = (Y_{1(1)}^0, \ldots, Y_{I(n_I)}^0)$ is treated as fixed, $q_{ij}$s are defined as in Theorem \ref{thm:perm_set}, and $A_{\bigcdot\bigcdot}$ follows the distribution in \eqref{eq:dist_A_set} for some $u_{ij}$s in $[0,1]$. 
\end{theorem}

The randomization distribution of treatment assignments $A_{\bigcdot\bigcdot}$ in \eqref{eq:dist_A_set} has the same form as that under Rosenbaum's sensitivity analysis framework \citep{GKR2000}. 
Following \citet{GKR2000} and \citet{WL23}, 
we are then able to conduct valid tests for $H_0$ with bounds on the 
maximum or quantiles of the $\Gamma_i$s.
For conciseness, 
we omit the detailed implementation here.  

\subsection{Adaptive permutation inference and its connection to  adaptive randomization inference}\label{sec:adapt_set}

Similar to \S \ref{sec:adapt_pair}, we can estimate the matching quality using external samples and conduct sensitivity analysis focusing more on the unmeasured confounding strength. Specifically, let $\hat{p}(\cdot \mid x)$ be the estimator for the conditional density of $Y^0$ given $X$ based on some external samples.  For all $ij$, we then estimate $p_{\text{m}ij}$  by $\hat{p}_{\text{m}ij}$, which has the same expression as $p_{\text{m}ij}$ but with $p(\cdot \mid \cdot)$ replaced by $\hat{p}(\cdot \mid \cdot)$. 
Let $\tilde{\Gamma}_{\text{u}i} = \Gamma_{\text{u}i} \cdot \max_{j,k\in [n_i]} \{ (p_{\text{m}ij}/\hat{p}_{\text{m}ij})/(p_{\text{m}ik}/\hat{p}_{\text{m}ik}) \}$. 
If the density estimation is precise, then 
$\tilde{\Gamma}_{\text{u}i} \approx \Gamma_{\text{u}i}$ measures mainly the unmeasured confounding strength  in \eqref{eq:confound_strength_set}, which reduces to $1$ if there is no unmeasured confounding. 
For $u_{ij}$s in $[0,1]$, 
define $A_{\bigcdot\bigcdot}$ such that
\begin{align}\label{eq:cov_adap_set}
\Pr(A_{\bigcdot\bigcdot} = a_{\bigcdot\bigcdot})
& = 
\prod_{i=1}^I \frac{\exp\{ \sum_{j=1}^{n_i} a_{ij}(\log \hat{p}_{\text{m}ij}+ \tilde{\gamma}_{\text{u}i} u_{ij})\}}{
	\sum_{j=1}^{n_i}\exp(\log \hat{p}_{\text{m}ij}+ \tilde{\gamma}_{\text{u}i} u_{ij})
}
\quad \textup{ with } 
\tilde{\gamma}_{\text{u}i} = \log \tilde{\Gamma}_{\text{u}i} \ge 0 \textup{ for all } i
\end{align}
if $a_{ij}$s are all in $\{0,1\}$ and $\sum_{j=1}^{n_i} a_{ij} = 1$ for all $i$, and $\Pr(A_{\bigcdot\bigcdot} = a_{\bigcdot\bigcdot}) = 0$ otherwise.

\begin{theorem}\label{thm:link_perm_rand_set_adap}
The permutation distribution of the test statistic in Theorem \ref{thm:perm_set}
is the same as the randomization distribution of 
$
t(A_{\bigcdot\bigcdot}, Y^0_{\bigcdot(\bigcdot)}) = 
\sum_{i=1}^I \sum_{j=1}^{n_i} A_{ij} q_{ij},
$
where $Y^0_{\bigcdot(\bigcdot)} = (Y_{1(1)}^0, 
\ldots, Y_{I(n_I)}^0)$ is treated as fixed, $q_{ij}$s are defined as in Theorem \ref{thm:perm_set},   
and $A_{\bigcdot\bigcdot}$ follows the distribution in \eqref{eq:cov_adap_set} for some $u_{ij}$s in $[0,1]$. 
\end{theorem}

From Theorem \ref{thm:link_perm_rand_set_adap}, we can then perform sensitivity analysis with a uniform bound on the $\tilde{\Gamma}_{\text{u}i}$s using the approach in \citet{pimentel2022covariate}. 
Despite the similarity in implementation, our permutation inference differs from their randomization inference in important aspects as discussed at the end of \S \ref{sec:adapt_pair}.

\section{Extensions}

\subsection{General sharp null hypotheses and test inversion}\label{sec:test_inversion}

We now extend the previous sensitivity analyses to general sharp null hypotheses. Consider, for example, the sharp null hypothesis of a certain constant effect $c\in \mathbb{R}$, which can be written as $H_c: Y^1 - Y^0 = c$ at the population level. 
Then for any unit we must have $Y^0 = Y - c Z$ under $H_c$, i.e., the control potential outcome can be imputed based on the observed outcome, observed assignment and the hypothesized effect. 
Consequently, the control potential outcomes are known under $H_c$ for all units in the observed data as well as the matched data. 
Thus, we can conduct sensitivity analyses using the same procedure developed in \S \ref{sec:pair}--\ref{sec:set}. 
In addition, as commented in \S \ref{sec:adapt_pair}, when estimating the conditional density of $Y^0$ given $X$ in the adaptive permutation inference, we need to impute $Y^0$ using the sharp null hypothesis as well.

We can also consider a sequence of sharp nulls with various constant effects, i.e., $H_c$ for various $c\in \mathbb{R}$, 
and use the standard test inversion to get interval estimates for the treatment effect. 
Specifically, if the treatment effects are indeed constant across all units, then the set of $c$ such that we are not able to reject $H_c$ at significance level $\alpha\in (0,1)$ forms a $1-\alpha$ confidence interval for the true constant treatment effect.

\subsection{Covariate adjustment}\label{sec:cov_adj}

\citet{Rosenbaum2002cov} and \citet[][Chapter 7]{rosenbaum2025introduction} suggested covariate adjustment in sensitivity analysis, 
aiming to improve inferential power by reducing variability in potential outcomes.
For example, he proposed to first conduct regression of the control potential outcomes $Y_{ij}^0$s on the covariates $X_{ij}$s and obtain residuals $\varepsilon_{ij}$s, and then construct test statistics based on the residuals instead of the original $Y_{ij}^0$s. 
Under the randomization-based inference framework, 
both the potential outcomes and covariates have been conditioned on, and they, as well as the corresponding residuals, can essentially be viewed as fixed constants. 
We can then perform Rosenbaum’s randomization-based sensitivity analysis for this covariate-adjusted test statistic in almost the same way as that for the original unadjusted one, except that the potential outcomes $Y_{ij}^0$ are now replaced by the residuals, or equivalently, the covariate-adjusted potential outcomes $\varepsilon_{ij}$s.  
However, such a procedure can be hard to justify and may not be valid 
under our permutation-based sensitivity analysis framework, because it uses the potential outcomes of the matched units, particularly the ordering of these potential outcomes besides their permutations.  

Below we discuss a way to perform covariate adjustment within the permutation-based framework using external samples. 
Specifically, 
we can fit a regression of the control potential outcomes on covariates using external samples, 
where the control potential outcomes are either observed or imputed based on the sharp null of interest. 
Denote the fitted regression function by $f(x)$. 
We can then perform the same permutation-based sensitivity analysis as in \S \ref{sec:pair}–\ref{sec:set}, but with the original control potential outcomes $Y^0_{ij}$ in the test statistic replaced by the covariate-adjusted potential outcomes $\tilde{Y}^0_{ij} = Y^0_{ij} - f(X_{ij})$, which are essentially residuals from the fitted regression. This approach is analogous to covariate adjustment proposed by Rosenbaum. 
A key distinction, however, is that Rosenbaum permits fitting the regression using the matched units, whereas we need to use external samples. 
The results in \S \ref{sec:pair}–\ref{sec:set} remain valid under covariate adjustment, except that the control potential outcomes are replaced by the covariate-adjusted potential outcomes.

We finally comment on an additional advantage of covariate adjustment, which is the potential to improve ``matching quality''. 
Consider a matched pair study as an example. 
If, after covariate adjustment, the adjusted potential outcome $\tilde{Y}^0 = Y^0 - f(X)$ becomes independent of the covariate $X$, then the matching quality, defined as in \eqref{eq:decomp_ps} with $Y^0$ replaced by $\tilde{Y}^0$, reduces to $1$, as in the case of exact matching.

\section{Numerical illustration}\label{sec:match_no_conf}

We now conduct numerical experiments to illustrate the proposed sensitivity analyses. 
Here we consider the case where matching is inexact and there is no unmeasured confounding; see the supplementary material for simulations in the presence of unmeasured confounding.
Specifically, similar to \citet[][Example 1]{Guo2022}, 
we generate data as i.i.d.~samples from the following model: 
\begin{align}\label{eq:model_no_unmeasured}
X \sim \text{Uniform}(0,1), 
\ \  
Y(1) = Y(0) \mid X \sim \mathcal{N}( X, 0.2^2), 
\ \  
\Pr\{ Z = 1 \mid X, Y(0) \} =  0.2 +  0.5 \cdot X,
\end{align}
under which $Z \ind (Y(1), Y(0)) \mid X$ and thus there is no unmeasured confounding. 
The study dataset consists of $N=1000$ i.i.d.~samples from 
\eqref{eq:model_no_unmeasured}. 
We conduct optimal pair matching without replacement based on the Mahalanobis distance in covariates \citep{Hansen2006}, where we match each treated unit with a control unit. 
We consider various sensitivity analyses for testing the sharp null hypothesis $H_0$ of no treatment effect, using the default m-statistic in \citet{senm17}.

\begin{figure}[htbp]
\centering
\includegraphics[width=0.7\linewidth]{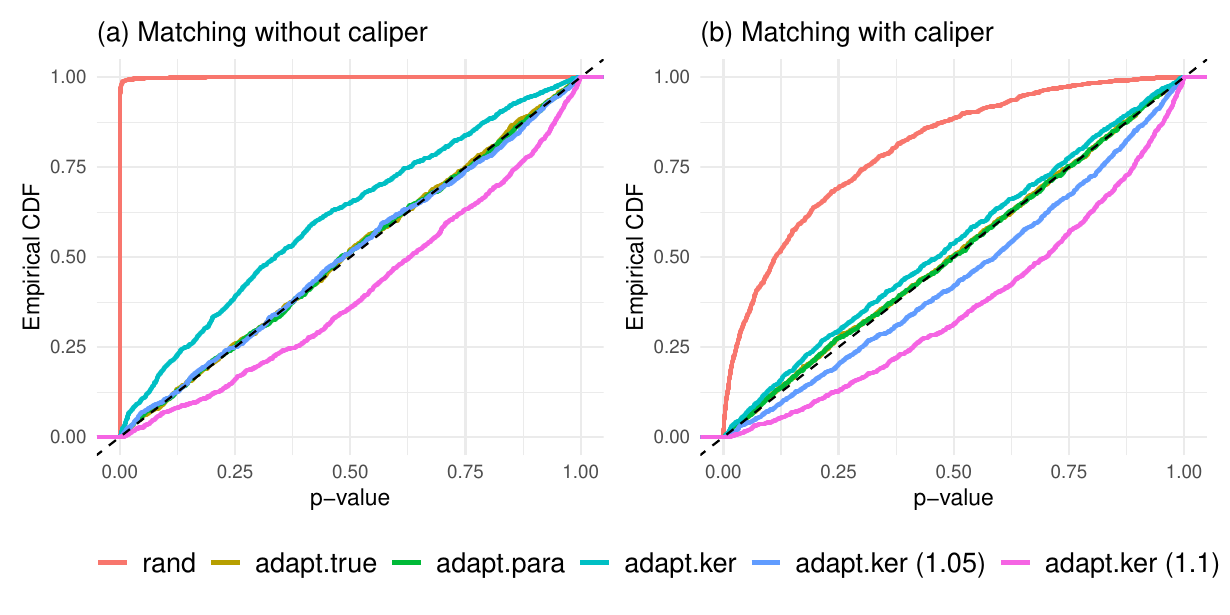}
\caption{Empirical distributions of $p$-values from various sensitivity analyses, under matching without caliper in (a) and with caliper in (b). The six curves in each plot show p-value distributions under six sensitivity analyses. ``rand'' assumes exact matching and no unmeasured confounding, and essentially conducts a randomization test. ``adapt.true'' assumes the true conditional distribution of $Y^0$ given $X$ is known, and conducts adaptive sensitivity analysis assuming no unmeasured confounding. ``adapt.para'' and ``adapt.ker'' estimate the conditional distribution of $Y^0$ given $X$ from external samples using either parametric or nonparametric approaches, and conduct adaptive sensitivity analysis assuming no unmeasured confounding. ``adapt.ker (1.05)'' and ``adapt.ker (1.1)'' use nonparametric density estimation and conduct adaptive sensitivity analysis assuming that $\tilde{\Gamma}_{\textup{u}i}$ is bounded by $1.05$ or $1.1$ for all $i$.
}
\label{fig:no_conf}
\end{figure}

First, we pretend that matching is exact and assume there is no unmeasured confounding. 
We can then test $H_0$ using uniform permutations within each matched pair. This is equivalent to the randomization test under a paired completely randomized experiment. 
The curve labeled ``rand'' in Fig.~\ref{fig:no_conf}(a) shows the empirical distribution of the p-value over $10^3$ simulated datasets from \eqref{eq:model_no_unmeasured}. 
Most of the p-values are about zero. 
Obviously,  the p-value is invalid with substantial type-I error inflation, which is coherent with the theoretical finding in \citet{Guo2022}. 
The invalidity of the randomization test is due to the inexactness of matching. 
Specifically,
recall the bias measure $\Gamma_i$s from \eqref{eq:decomp_ps}.
Across the $10^3$ simulations, the maximum of the biases from all matched pairs ranges from $7.57$ to $9.79 \times 10^{12}$, with a median of $1.52 \times 10^5$, 
and the $80\%$ quantile of the biases ranges from $1.17$ to $21.34$, with a median of $2.68$.

Second, we conduct a sensitivity analysis imposing bounds on either the maximum or the $80\%$ quantile  of the biases $\Gamma_i$, where the bound is set to the corresponding true value to ensure correct specification.
The resulting p-values are valid but highly conservative, all being approximately $1$. This is because the biases vary across matched pairs, with many falling well below the imposed bounds.

Third, we consider adaptive sensitivity analysis in \S \ref{sec:adapt_pair}, assuming that we know the true conditional distribution of the potential outcome $Y^0$ given covariate $X$. 
Recall the notation in  \S \ref{sec:adapt_pair}. 
We know the true values of the $\ratio_{\text{m}i}$s. 
We then conduct the adaptive sensitivity analysis with $\hat{\ratio}_{\text{m}i}$ set to $\ratio_{\text{m}i}$ for all $i$, under the constraint that $\tilde{\Gamma}_{\textup{u}i} =  1$ for all $i$. 
The constraint here holds because there is no unmeasured confounding under \eqref{eq:model_no_unmeasured}. 
The curve labeled ``adapt.true'' in Fig.~\ref{fig:no_conf}(a) shows the empirical distribution of the p-value over $10^3$ simulated datasets from \eqref{eq:model_no_unmeasured}. 
We can see that the p-value is approximately uniformly distributed on $(0,1)$, indicating its validity and exactness. 

Fourth, we consider the adaptive sensitivity analysis in \S \ref{sec:adapt_pair}, using an external independent dataset of size 1000 to estimate the conditional density of $Y^0$ given $X$.
We consider both parametric and nonparametric approaches for estimating the conditional density: one based on a correctly specified linear regression model with homoscedastic Gaussian error, and the other based on kernel density estimation \citep{ks}.
We then perform the adaptive sensitivity analysis with $\hat{\ratio}_{\text{m}i}$s calculated based on the estimated density of $Y^0$ given $X$, under the constraint that $\tilde{\Gamma}_{\textup{u}i} =  1$ for all $i$. 
In this case, the constraint holds only approximately, due to the uncertainty in the estimation of $\ratio_{\text{m}i}$s. 
The curves labeled ``adapt.para'' and ``adapt.ker'' in Fig.~\ref{fig:no_conf}(a)  show the empirical distributions of the p-value over $10^3$ simulated datasets from \eqref{eq:model_no_unmeasured}, under the two approaches for density estimation. 
The p-value is approximately uniformly distributed on $(0,1)$ when using the correctly specified parametric model for density estimation. 
In contrast, using the nonparametric kernel method introduces moderate size distortion due to the error in density estimation.

To accommodate errors in density estimation, which can be challenging as the covariate dimension increases, we also consider the adaptive sensitivity analysis under the constraint that $\tilde{\Gamma}_{\textup{u}i}$ is bounded by $1.05$ or $1.1$ for all $i$. Ideally, $\tilde{\Gamma}_{\textup{u}i}$s should be $1$ without unmeasured confounding, but we allow slightly looser constraints to mitigate potential inaccuracy in density estimation. The resulting empirical distributions of the p-values are shown by the remaining two curves in Fig.~\ref{fig:no_conf}, labeled as ``adapt.ker (1.05)'' and ``adapt.ker (1.1)’’. With this slight relaxation, the p-value becomes approximately valid or slightly conservative.

We also consider improving the quality of matching through caliper \citep{Rosenbaum01021985}. 
Specifically, we discard matched pairs whose absolute difference in the covariate is greater than a quarter of its pooled standard deviation. 
The results of the above sensitivity analyses are analogously shown in Fig.~\ref{fig:no_conf}(b). 
Compared to Fig.~\ref{fig:no_conf}(a),
the p-value based on kernel density estimation exhibits less severe size distortion, suggesting that improved matching quality can help mitigate inaccuracies in density estimation. 
Recall that the adaptive sensitivity analysis relies essentially on conditional density ratios between covariate values within the same matched set. Improved matching can make covariates more similar within each matched set, thereby reducing the sensitivity of the ratio estimates. Moreover, as the analysis depends only on these ratios, it is of interest to explore more robust and flexible estimation strategies.

In the supplementary material, we further present simulations for testing a sequence of constant treatment effects and the corresponding test inversion for interval estimation, as discussed in \S \ref{sec:test_inversion}.

\section{Discussion}%
In the paper we focus on sharp nulls. It will be interesting to extend it to weak nulls on average effects or effect quantiles \citep{Fogarty2020student, CDLM21quantile, SL22quantile, ZL24quantile}.

\section*{Acknowledgment}
We thank the editor, the associate editor and the reviewers for insightful and constructive comments. Li was partly supported by the U.S. National Science Foundation (DMS-2400961).

\section*{Supplementary material}
The supplementary material includes (i) simplified bias measures in special cases, (ii) additional numerical experiments, 
and (iii) all the technical details, including the proofs of all the theorems.

\bibliographystyle{plainnat}
\bibliography{reference}

\newpage
\setcounter{equation}{0}
\setcounter{section}{0}
\setcounter{figure}{0}
\setcounter{example}{0}
\setcounter{proposition}{0}
\setcounter{corollary}{0}
\setcounter{theorem}{0}
\setcounter{lemma}{0}
\setcounter{table}{0}
\setcounter{condition}{0}

\renewcommand {\theproposition} {A\arabic{proposition}}
\renewcommand {\theexample} {A\arabic{example}}
\renewcommand {\thefigure} {A\arabic{figure}}
\renewcommand {\thetable} {A\arabic{table}}
\renewcommand {\theequation} {A\arabic{equation}}
\renewcommand {\thelemma} {A\arabic{lemma}}
\renewcommand {\thesection} {A\arabic{section}}
\renewcommand {\thetheorem} {A\arabic{theorem}}
\renewcommand {\thecorollary} {A\arabic{corollary}}
\renewcommand {\thecondition} {A\arabic{condition}}
\renewcommand {\thermk} {A\arabic{rmk}}

\renewcommand {\thepage} {A\arabic{page}}
\setcounter{page}{1}

\begin{center}
	\bf \LARGE 
	Supplementary Material 
\end{center}

\section{Bias measures in special cases}\label{sec:simp_special}

In this section, we simplify the bias measures in the cases of exact matching or no unmeasured confounding. 
Let $p_{z}(\cdot \mid x) \equiv p(\cdot \mid x, z)$ denote the density for the conditional distribution of $Y^\control$ given $X=x$ and $Z=z$, for $z=0,1$. 

We consider first the matched pair studies. 
From \S \ref{sec:bias_pair} and \eqref{eq:r_i_pair_deri} and \eqref{eq:r_i_pair_deri_equiv} in the later \S \ref{sec:tech_detail_pair}, for all $i$, $\ratio_i$ in \eqref{eq:decomp_ps} has the following equivalent forms:
\begin{align}\label{eq:ratio_pair_two_equiv}
	\ratio_i = \frac{p_1( Y^0_{i(2)} \mid X_{i1}) p_0( Y^0_{i(1)} \mid X_{i2})}{
		p_1( Y^0_{i(1)} \mid X_{i1}) p_0( Y^0_{i(2)} \mid X_{i2})
	}
	= 
	\frac{
		p(Y_{i(2)}^0 \mid X_{i1}) / p(Y_{i(2)}^0 \mid X_{i2})
	}{
		p(Y_{i(1)}^0 \mid X_{i1}) / p(Y_{i(1)}^0 \mid X_{i2})
	}
	\frac{
		\pi_0(X_{i2}, Y_{i(1)}^0)/\pi_0(X_{i2}, Y_{i(2)}^0)
	}{
		\pi_1(X_{i1}, Y_{i(1)}^0)/\pi_1(X_{i1}, Y_{i(2)}^0)
	}. 
\end{align}

\begin{example}[Exact matching in matched pair studies]\label{eg:exact_match}
	Suppose that matching is exact. 
	In this case, 
	let $\tilde{X}_{i} = X_{i1} = X_{i2}$ denote the common covariate value for units in matched pair $i\in [I]$. We can then simplify $\ratio_i$ in \eqref{eq:ratio_pair_two_equiv} as
	\begin{align}\label{eq:ri_exact_match}
		\ratio_i & = \frac{p_1( Y^0_{i(2)} \mid \tilde{X}_{i})/ p_0( Y^0_{i(2)} \mid \tilde{X}_{i})}{
			p_1( Y^0_{i(1)} \mid \tilde{X}_{i}) / p_0( Y^0_{i(1)} \mid \tilde{X}_{i})
		}
		= 
		\frac{
			\pi_1(\tilde{X}_{i}, Y_{i(2)}^0)/\pi_0(\tilde{X}_{i}, Y_{i(2)}^0)
		}{
			\pi_1(\tilde{X}_{i}, Y_{i(1)}^0)/\pi_0(\tilde{X}_{i}, Y_{i(1)}^0)
		}
		.  
	\end{align}
	In \eqref{eq:ri_exact_match}, 
	$\ratio_{i}$ has two equivalent forms. 
	The first one involves contrasts of potential outcome distributions between treated and control units, 
	which have been considered in the sensitivity analysis by, e.g, \citet{Franks2020}.  
	The second one is the odds ratio of the treatment probabilities of the two matched units. 
	The corresponding $\Gamma_i$ then simplifies to 
	\begin{align*}
		\Gamma_i & = \max\{ \ratio_{\text{u}i}, \ratio_{\text{u}i}^{-1} \}
		= \max\left\{ \frac{
			\pi_1(\tilde{X}_{i}, Y_{i(2)}^0)/\pi_0(\tilde{X}_{i}, Y_{i(2)}^0)
		}{
			\pi_1(\tilde{X}_{i}, Y_{i(1)}^0)/\pi_0(\tilde{X}_{i}, Y_{i(1)}^0)
		}, \ \ 
		\frac{
			\pi_1(\tilde{X}_{i}, Y_{i(1)}^0)/\pi_0(\tilde{X}_{i}, Y_{i(1)}^0)
		}{
			\pi_1(\tilde{X}_{i}, Y_{i(2)}^0)/\pi_0(\tilde{X}_{i}, Y_{i(2)}^0)
		}
		\right\}\\
		& = 
		\max\left\{ \frac{
			\pi_1(\tilde{X}_{i}, Y_{i2}^0)/\pi_0(\tilde{X}_{i}, Y_{i2}^0)
		}{
			\pi_1(\tilde{X}_{i}, Y_{i1}^0)/\pi_0(\tilde{X}_{i}, Y_{i1}^0)
		}, 
		\ \ 
		\frac{
			\pi_1(\tilde{X}_{i}, Y_{i1}^0)/\pi_0(\tilde{X}_{i}, Y_{i1}^0)
		}{
			\pi_1(\tilde{X}_{i}, Y_{i2}^0)/\pi_0(\tilde{X}_{i}, Y_{i2}^0)
		}
		\right\},
	\end{align*}
	which essentially reduces to \citet{Rosenbaum02a}'s bias measure, with the unmeasured confounding there being the control potential outcomes.
\end{example}

\begin{example}[No unmeasured confounding in matched pair studies]\label{eg:no_unmeasured}
	Suppose that there is no unmeasured confounding. 
	In this case, $\ratio_i$ in \eqref{eq:ratio_pair_two_equiv} simplifies to 
	\begin{align}\label{eq:ratio_pair_no_conf}
		\ratio_{i} 
		& = 
		\frac{p_1( Y^0_{i(2)} \mid X_{i1}) p_0( Y^0_{i(1)} \mid X_{i2})}{
			p_1( Y^0_{i(1)} \mid X_{i1}) p_0( Y^0_{i(2)} \mid X_{i2})
		}   
		= 
		\frac{
			p(Y_{i(2)}^0 \mid X_{i1}) / p(Y_{i(2)}^0 \mid X_{i2})
		}{
			p(Y_{i(1)}^0 \mid X_{i1}) / p(Y_{i(1)}^0 \mid X_{i2})
		}
		, 
	\end{align}
	where the equivalence between the two expressions is obvious since, without unmeasured confounding, $p(\cdot\mid x) = p_1(\cdot\mid x) = p_0(\cdot\mid x)$ for any $x$. 
	In this case, the magnitude of the $\Gamma_i$s are mainly driven by the within-pair covariate imbalance, which are generally sought to be minimized during matching.
\end{example}

We consider then general matched set studies. 
As demonstrated in the later \S \ref{sec:tech_set}, 
the biases from uniform permutation can be bounded by 
$\bar{\Gamma}_i \equiv \max_{1\le j \ne k \le n_i, 2\le l \le n_i} \ratio_{ijkl}$ for $i \in [I]$, 
where $\ratio_{ijkl}$ is defined in the later \eqref{eq:ratio_p_pi_sigma} and has the following form:  
\begin{align*}
	\ratio_{ijkl} = \frac{
		p(Y^0_{i(j)} \mid X_{i1})/p(Y^0_{i(j)} \mid X_{il})
		\cdot
		\pi_0(X_{il}, Y^0_{i(k)} )/\pi_0(X_{il}, Y^0_{i(j)} ) 
	}{
		p(Y^0_{i(k)} \mid X_{i1})/p(Y^0_{i(k)} \mid X_{il})
		\cdot
		\pi_1(X_{i1}, Y^0_{i(k)}) / \pi_1(X_{i1}, Y^0_{i(j)})
	}. 
\end{align*}
Obviously, $\ratio_{ijkl}$ has a similar form as that in \eqref{eq:ratio_pair_two_equiv}, 
and can be similarly simplified in the case of exact matching or no unmeasured confounding, as in Examples \ref{eg:exact_match} and \ref{eg:no_unmeasured}. 
For example, when matching is exact, 
denoting the common covariate value in matched set $i$ by $\tilde{X}_i$,
we can simplify $\bar{\Gamma}_i$ as
\begin{align*}
	\bar{\Gamma}_i 
	& \equiv \max_{1\le j \ne k \le n_i, 2\le l \le n_i}   
	\ratio_{ijkl} = 
	\max_{1\le j \ne k \le n_i}\frac{
		\pi_1(\tilde{X}_{i}, Y^0_{i(j)}) /\pi_0(\tilde{X}_{i}, Y^0_{i(j)} ) 
	}{
		\pi_1(\tilde{X}_{i}, Y^0_{i(k)}) /\pi_0(\tilde{X}_{i}, Y^0_{i(k)} )
	}
	\\
	& = 
	\max_{1\le j \ne k \le n_i}\frac{
		\pi_1(\tilde{X}_{i}, Y^0_{ij}) /\pi_0(\tilde{X}_{i}, Y^0_{ij} ) 
	}{
		\pi_1(\tilde{X}_{i}, Y^0_{ik}) /\pi_0(\tilde{X}_{i}, Y^0_{ik} )
	}, 
\end{align*}
which is the maximum odds ratio of the treatment probabilities for units within the same matched set. 
This is the same as Rosenbaum's $\Gamma$ for measuring biases from uniform randomization of treatment assignment within each matched set, with unmeasured confounding there being the control potential outcomes.

\section{Numerical experiments}

This section presents additional numerical experiments illustrating the proposed sensitivity analyses.

\subsection{Inexact matching and unmeasured confounding}\label{sec:inexact_conf}

Analogous to \S \ref{sec:match_no_conf}, we now consider the case where matching is inexact and there is  unmeasured confounding.
We consider i.i.d.~samples from the following model:
\begin{align}\label{eq:model_conf}
	X & \sim \text{Uniform}(0,1), \nonumber\\
	Y(1) = Y(0) \mid X & \sim \mathcal{N}( 2X, 0.2^2), \nonumber\\
	\text{logit} \big( \Pr\{ Z = 1 \mid X, Y(0) \} \big)  & = -1 + 0.5 \cdot X + 0.5 \cdot Y(0),
\end{align}
under which $Z \nind (Y(1), Y(0)) \mid X$ and thus there is unmeasured confounding. 
We consider the same simulation as in \S \ref{sec:match_no_conf} with optimal pair matching, except that the study dataset consists of $N=1000$ i.i.d.~samples from the model in \eqref{eq:model_conf}. 
We consider again testing the sharp null hypothesis $H_0$ of no treatment effect and conduct sensitivity analysis using the approaches described below. We use the same test statistic as in \S \ref{sec:match_no_conf}. 
Recall the notation in \S \ref{sec:pair}. 
For each matched dataset, 
we know the true values of the $\ratio_{\text{u}i}$s. 
We then define $\Gamma_{\text{u}i} = \max\{\ratio_{\text{u}i}, \ratio_{\text{u}i}^{-1}\}$ for all $i$ to denote the strength of unmeasured confounding.

\begin{figure}
	\centering
	\includegraphics[width=0.85\linewidth]{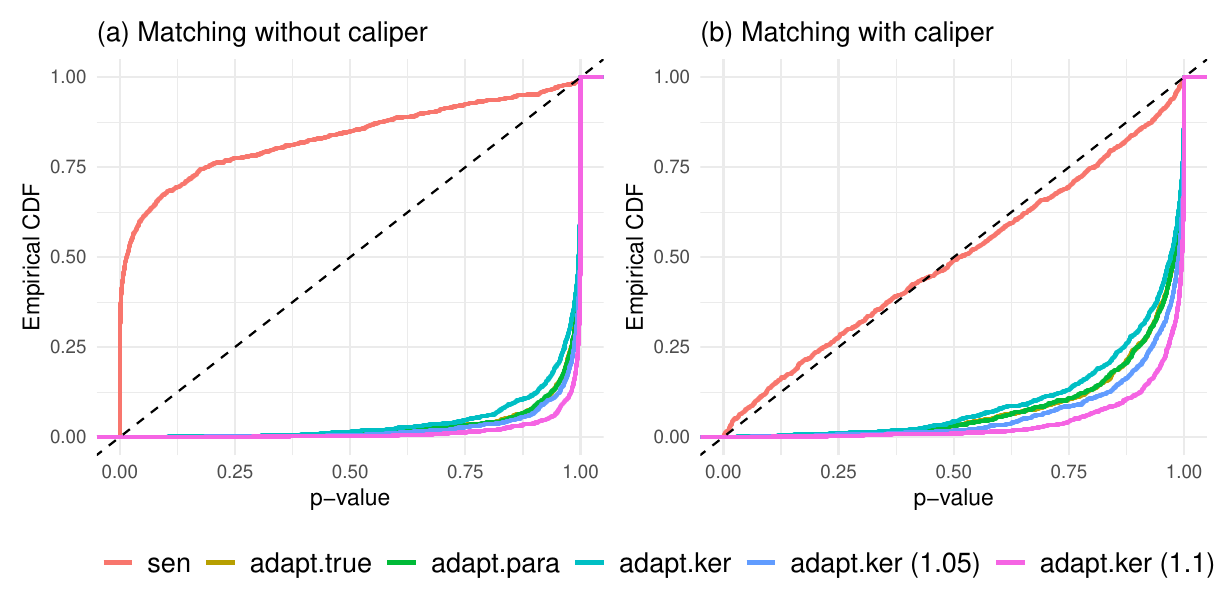}
	\caption{
		Empirical distributions of $p$-values from various sensitivity analyses, under matching without caliper in (a) and with caliper in (b). The six curves in each plot show p-value distributions under six sensitivity analyses. ``sen'' assumes the biases $\Gamma_i$s are uniformly bounded by  $\max_{i\in [I]}\Gamma_{\text{u}i}$, the maximum strength of unmeasured confounding across all matched pairs. ``adapt.true'' assumes the true conditional distribution of $Y^0$ given $X$ is known, and conducts adaptive sensitivity analysis assuming $\tilde{\Gamma}_{\text{u}i}$s are uniformly bounded by $\max_{i\in [I]}\Gamma_{\text{u}i}$. ``adapt.para'' and ``adapt.ker'' estimate the conditional distribution of $Y^0$ given $X$ from external samples using either parametric or nonparametric approaches, and conduct adaptive sensitivity analysis assuming $\tilde{\Gamma}_{\text{u}i}$s are uniformly bounded by $\max_{i\in [I]}\Gamma_{\text{u}i}$. ``adapt.ker (1.05)'' and ``adapt.ker (1.1)'' use nonparametric density estimation and conduct adaptive sensitivity analysis assuming that $\tilde{\Gamma}_{\textup{u}i}$ is bounded by $1.05 \times \max_{i\in [I]}\Gamma_{\text{u}i}$ or $1.1 \times \max_{i\in [I]}\Gamma_{\text{u}i}$ for all $i$.
	}
	\label{fig:conf}
\end{figure}

\begin{figure}
	\centering
	\includegraphics[width=0.85\linewidth]{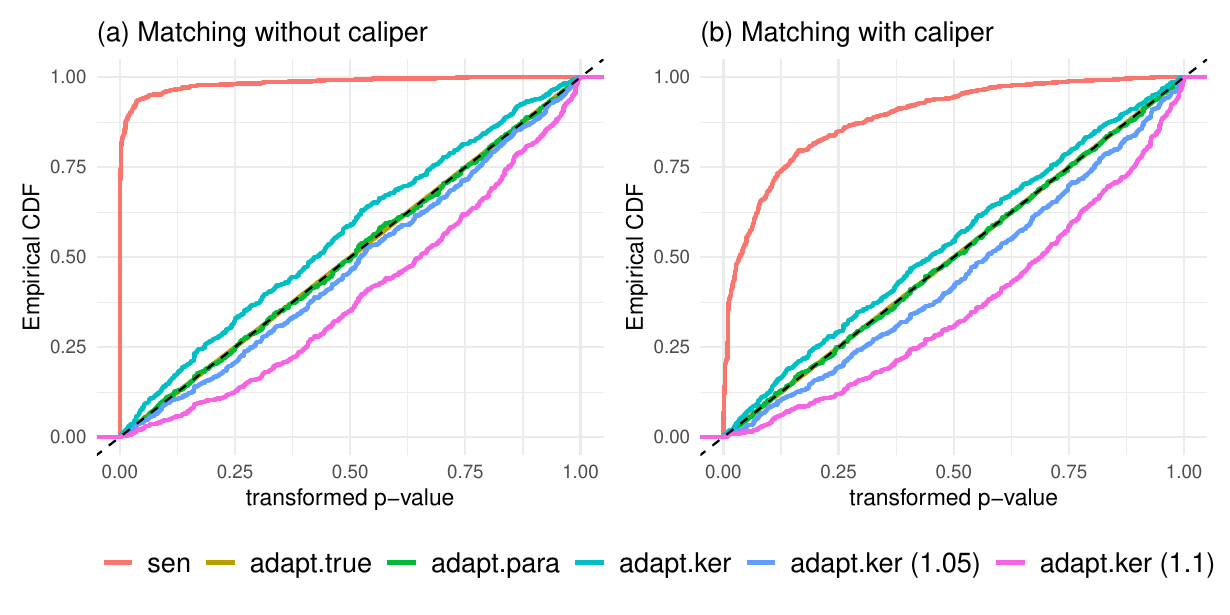}
	\caption{
		Empirical distributions of transformed $p$-values from various sensitivity analyses, under matching without caliper in (a) and with caliper in (b).
		The p-values are those in Fig. \ref{fig:conf}, and they are transformed using the empirical distribution function of the p-value from the corresponding adaptive sensitivity analysis based on the true density of $Y^0$ given $X$.
	}
	\label{fig:conf_transform}
\end{figure}

First, we 
conduct 
sensitivity analysis in \S \ref{sec:sen_base_unif_perm} under the constraint that the $\Gamma_i$s, which measure biases from uniform permutations,  are all bounded by $\max_{i\in [I]}\Gamma_{\text{u}i}$. 
If matching is exact, then this constraint will hold; otherwise, this constraint may fail. 
The curve labeled ``sen'' in 
Fig. \ref{fig:conf}(a) shows the empirical distribution of the p-value across $10^3$ simulations. Obviously, the p-value does not control the type-I error well and is thus not valid. 
This is not surprising since matching here is inexact.   

Second, we consider adaptive sensitivity analysis in \S \ref{sec:adapt_pair}, assuming we know the true conditional density of the potential outcome $Y^0$ given covariate $X$. 
In this case, we know the true values of the $\ratio_{\text{m}i}$s.
We conduct the adaptive sensitivity analysis with $\hat{\ratio}_{\text{m}i} = \ratio_{\text{m}i}$ for all $i$, under the constraint that $\tilde{\Gamma}_{\text{u}i}$s are uniformly bounded by $\max_{i\in [I]}\Gamma_{\text{u}i}$. 
This constraint must hold because $\tilde{\Gamma}_{\text{u}i}=\Gamma_{\text{u}i}$ for all $i$ by definition. 
The curve labeled ``adapt.true'' in 
Fig. \ref{fig:conf}(a) shows the empirical distribution of the p-value across all simulations. 
The p-value is valid, coherent with our theory, although conservative due to the heterogeneity of the unmeasured confounding strength across matched pairs. 

Third, we consider adaptive sensitivity analysis in \S \ref{sec:adapt_pair} without knowing the conditional distribution of $Y^0$ given $X$. 
We estimate the conditional density of $Y^0$ given $X$ using an external independent dataset of size $1000$. 
Similar to \S \ref{sec:match_no_conf}, 
we consider both parametric or nonparametric approaches for density estimation: one based on a correctly specified  linear regression model with homoscedastic Gaussian error, and the other based on kernel density estimation \citep{ks}. 
We then conduct adaptive sensitivity analysis with $\hat{\ratio}_{\text{m}i}$s calculated based on the estimated density, under the constraint that $\tilde{\Gamma}_{\text{u}i}$s are uniformly bounded by $\max_{i\in [I]}\Gamma_{\text{u}i}$.  
Here 
the constraint holds only approximately due to the uncertainty in the estimation of the conditional density of $Y^0$ given $X$. 
The curves labeled ``adapt.para'' and  ``adapt.ker'' in 
Fig. \ref{fig:conf}(a) show the empirical distributions of the p-value across all simulations under these two approaches for density estimation, respectively. 
The distribution of the p-value under the parametric density estimation is close to, and almost overlaps with, that based on the true density of $Y^0$ given $X$, 
while the distribution of the p-value under the nonparametric density estimation is less conservative than that based on the true density. 
We also consider slightly looser constraints to mitigate potential inaccuracy in density estimation, assuming that $\tilde{\Gamma}_{\text{u}i}$s are uniformly bounded by $1.05 \times \max_{i\in [I]}\Gamma_{\text{u}i}$ or $1.1 \times \max_{i\in [I]}\Gamma_{\text{u}i}$.  
The resulting distributions of the p-values are shown by the remaining two curves in Fig. \ref{fig:conf}(a), labeled ``adapt.ker (1.05)'' and ``adapt.ker (1.1)''. 
They are closer to, or slightly more conservative than, the p-value based on the true density of $Y^0$ given $X$.

To better visualize differences in the p-value distributions across various sensitivity analyses, we transform all p-values using the empirical distribution function of the p-value from the adaptive sensitivity analysis based on the true density of $Y^0$ given $X$. The resulting empirical distributions of the transformed p-values are shown in Fig. \ref{fig:conf_transform}(a). 
The transformed p-value under the adaptive sensitivity analysis with true density becomes uniformly distributed on $(0,1)$, while those using estimated densities or slightly looser constraints are either close to uniform, less conservative, or more conservative, as discussed earlier.

We also perform matching with caliper, where we discard matched pairs whose absolute difference in the covariate is greater than a quarter of its pooled standard deviation. 
The simulation results are analogously shown in Fig. \ref{fig:conf}(b) and \ref{fig:conf_transform}(b). 
The improved matching quality helps mitigate the anti-conservativeness from inaccuracies in nonparametric density estimation. 

\subsection{Test inversion}\label{sec:test_invert}
We now conduct two simulations to illustrate the test inversion for interval estimation of treatment effects under our proposed sensitivity analysis. 
The data generating models are similar to \eqref{eq:model_no_unmeasured} and \eqref{eq:model_conf}, except that the treatment effects are nonzero. 
The sensitivity analyses below focus mainly on the strength of unmeasured confounding.

\begin{figure}[htb]
	\centering
	\begin{subfigure}{0.5\textwidth}
		\centering
		\includegraphics[width=0.9\textwidth]{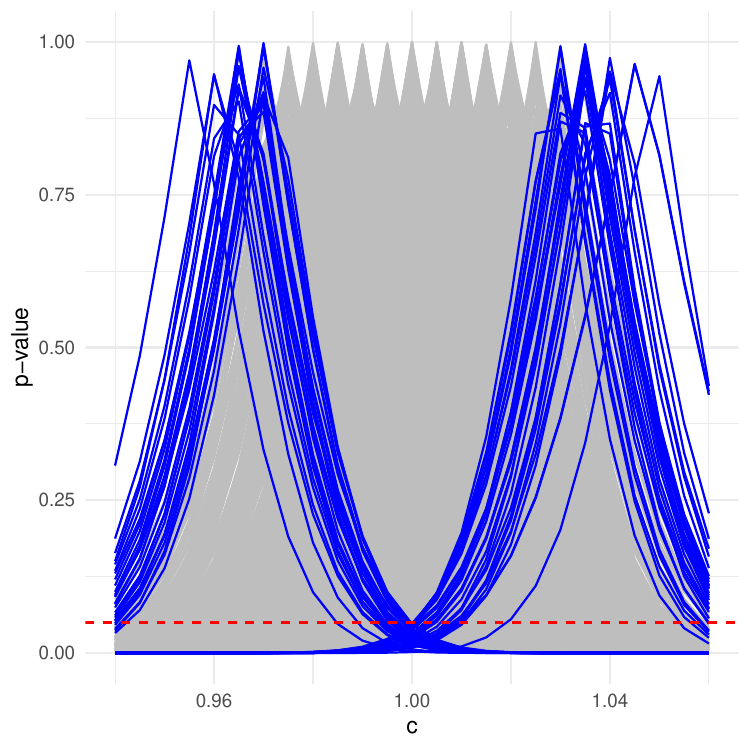}
		\caption{\centering }
	\end{subfigure}%
	\begin{subfigure}{0.5\textwidth}
		\centering
		\includegraphics[width=0.9\textwidth]{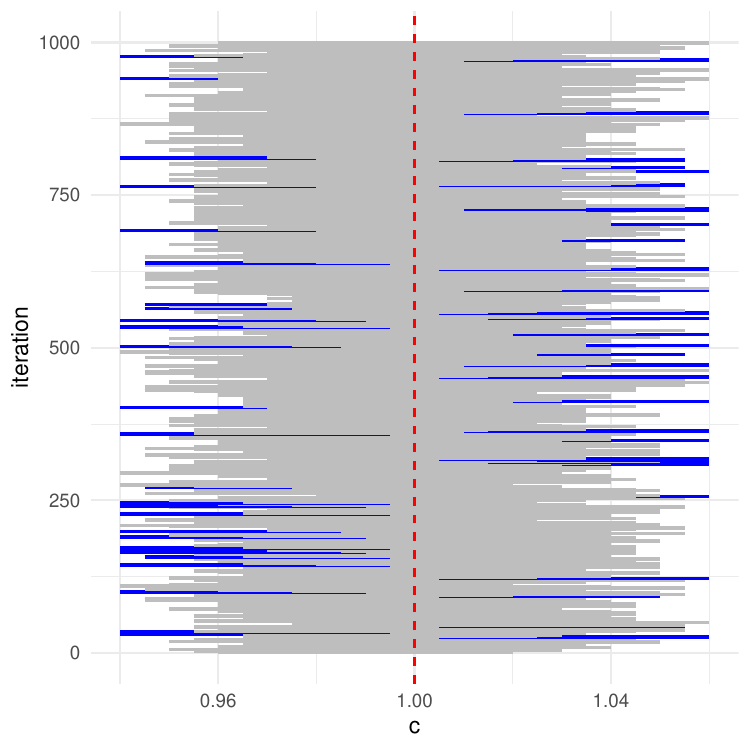}
		\caption{\centering }
	\end{subfigure}
	\caption{
		(a) shows the p-values from the adaptive sensitivity analysis at various $c$ across $10^3$ simulations, with data generated from the model in \eqref{eq:model_no_unmeasured_constant_1}. In (a), the blue lines indicate the simulations for which the p-value for the null at true treatment effect $c=1$ is less than or equal to $0.05$, 
		and the red horizontal line indicates the significance level $0.05$. 
		(b) shows the interval estimates from the standard test inversion across $10^3$ simulations. In (b), the blue lines indicate the simulations for which the interval does not cover the true constant treatment effect $c=1$, and the red vertical line indicates the true constant treatment effect $c=1$. 
	} 
	\label{fig:ci_no_conf}
\end{figure}

We first consider study data consisting of $N=1000$ i.i.d.~samples from the following model:
\begin{align}\label{eq:model_no_unmeasured_constant_1}
	X & \sim \text{Uniform}(0,1), \nonumber\\
	Y(0) \mid X & \sim \mathcal{N}( X, 0.2^2), 
	\nonumber
	\\
	Y(1) & = Y(0) + 1, 
	\nonumber\\
	\Pr\{ Z = 1 \mid X, Y(0) \} & =  0.2 +  0.5 \cdot X,
\end{align}
under which there is no unmeasured confounding, 
and conduct optimal pair matching. 
For any sharp null hypothesis $H_c: Y(1) - Y(0) = c$ at any given $c\in \mathbb{R}$, 
we conduct the adaptive sensitivity analysis in \S \ref{sec:adapt_pair}. 
We estimate the conditional density of $Y^0$ given $X$
based on a correctly specified  linear regression model with homoscedastic Gaussian error and an external independent dataset of size $1000$;  
note that we need to first impute the control potential outcomes based on the observed data and the null hypothesis $H_c$, in order to estimate the conditional density of $Y^0$ given $X$. 
We then obtain $\hat{R}_{\text{m}i}$s based on the estimated density and perform the adaptive sensitivity analysis under the constraint that $\tilde{\Gamma}_{\textup{u}i}=1$ for all $i$. 
Note that, 
when $c=1$, for which the null hypothesis $H_c$ holds under the model in \eqref{eq:model_no_unmeasured_constant_1}, this constraint holds approximately, up to some estimation error for the conditional density of $Y^0$ given $X$. 
Fig. \ref{fig:ci_no_conf}(a) shows the p-values at various $c$ across $10^3$ simulations, 
and Fig. \ref{fig:ci_no_conf}(b) shows the interval estimates across $10^3$ simulations at significance level $5\%$. 
From Fig. \ref{fig:ci_no_conf}, across the $10^3$ simulations, the interval covers the true constant treatment effect $94.3\%$ of the time. This confirms the validity of using test inversion to construct confidence intervals for the true treatment effects.

\begin{figure}[htbp]
	\centering
	\begin{subfigure}{0.5\textwidth}
		\centering
		\includegraphics[width=0.8\textwidth]{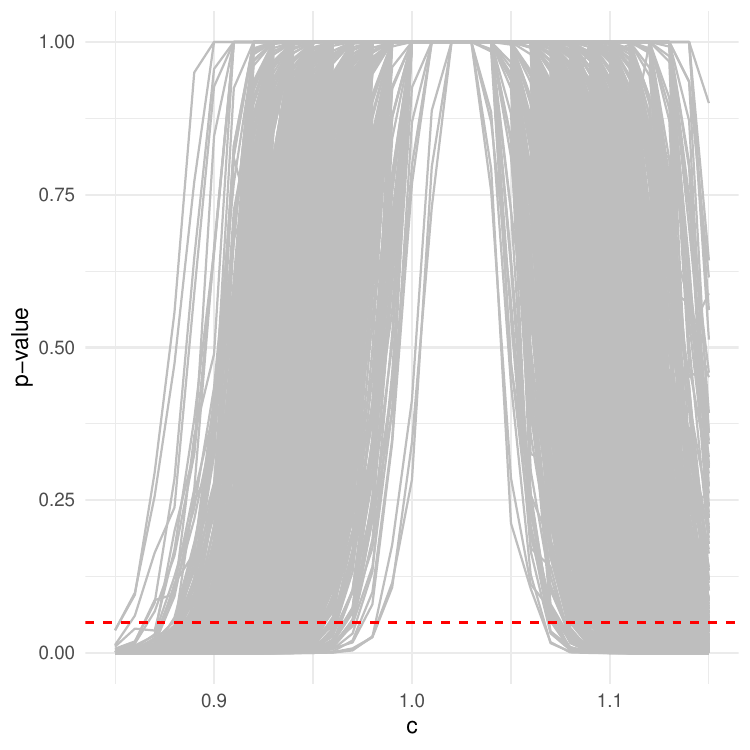}
		\caption{\centering }
	\end{subfigure}%
	\begin{subfigure}{0.5\textwidth}
		\centering
		\includegraphics[width=0.8\textwidth]{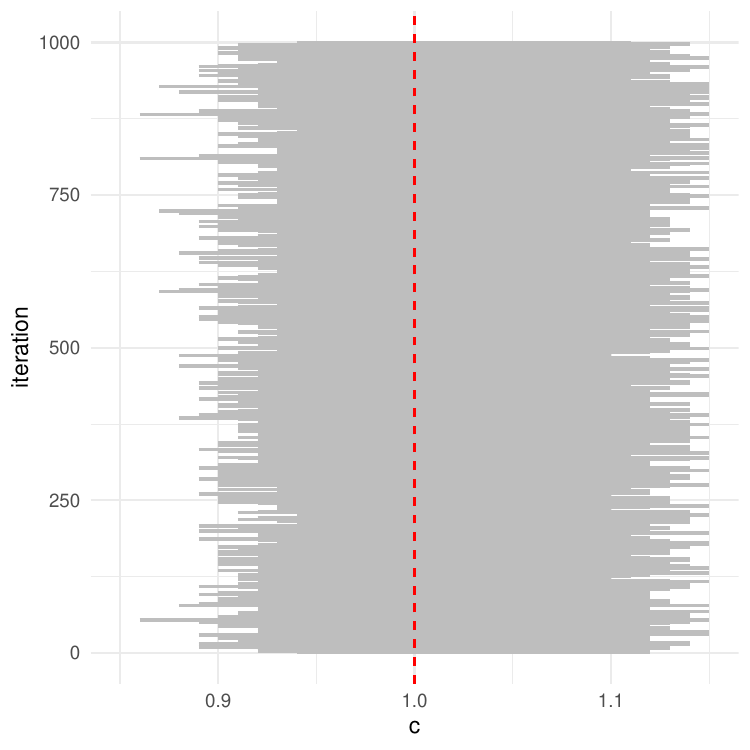}
		\caption{\centering }
	\end{subfigure}
	\caption{
		(a) shows the p-values from the adaptive sensitivity analysis at various $c$ across $10^3$ simulations, with data generated from the model in \eqref{eq:model_conf_constant_1}. 
		(b) shows the interval estimates from the standard test inversion across $10^3$ simulations.
		In (a),  the red horizontal line indicates the significance level $0.05$. 
		In (b), the red vertical line indicates the true constant treatment effect $c=1$. 
	} 
	\label{fig:ci_conf}
\end{figure}

We then consider study data consisting of i.i.d.~samples from the following model:
\begin{align}\label{eq:model_conf_constant_1}
	X & \sim \text{Uniform}(0,1), \nonumber\\
	Y(0) \mid X & \sim \mathcal{N}( 2X, 0.2^2), \nonumber\\
	Y(1) & = Y(0)+1, \nonumber
	\\
	\text{logit} \big(  \Pr\{ Z = 1 \mid X, Y(0) \} \big) & =  -1 + 0.5 \cdot X + 0.5 \cdot Y(0),
\end{align}
where there is unmeasured confounding. 
For any sharp null hypothesis $H_c: Y(1) - Y(0)=c$ at any given $c\in \mathbb{R}$, 
we conduct adaptive sensitivity analysis with $\hat{R}_{\text{m}i}$s obtained  from parametric density estimation using external samples; this is analogous to the previous case. 
We then consider the constraint that $\tilde{\Gamma}_i$s are uniformly bounded by $\max_{i\in [I]} \Gamma_{\textup{u}i}$, where $\Gamma_{\textup{u}i}$ denotes the true strength of unmeasured confounding as in \S \ref{sec:inexact_conf}. 
Again, when $c=1$, this constraint holds approximately, up to some estimation error for the conditional density of $Y^0$ given $X$. 
Fig. \ref{fig:ci_conf}(a) shows the p-values at various $c$ across $10^3$ simulations, 
and Fig. \ref{fig:ci_conf}(b) shows the interval estimates across $10^3$ simulations at significance level $5\%$. 
From Figure \ref{fig:ci_conf}, across the $10^3$ simulations, the interval covers the true constant treatment effect all the time. This conservativeness is due to the heterogeneity of unmeasured confounding strength across matched pairs; specifically, the unmeasured confounding strength for some pairs are strictly less than the imposed bound.

\section{Technical details for matched pair studies}
\subsection{Technical details for \S \ref{sec:bias_pair}}\label{sec:tech_detail_pair}

Below we study the conditional distribution of the potential outcomes $Y_{i\bigcdot}^0$s given $\mathcal{F}_{\text{pair}}$. 

First, we show that, given $\mathcal{F}_{\text{pair}}$, the control potential outcomes are mutually independent across all matched pairs. 
Note that (i) the matching algorithm depends only on the observed treatment assignments and the observed covariates for all the $N$ units prior to matching, and (ii) the matching structure, 
which specifies what unit corresponds to each matched unit in the matched data, 
are determined once given the observed assignments, observed covariates and the matching algorithm. 
From Assumption \ref{asmp:iid},  we can know that, conditional on 
$\mathcal{G}_{\text{pair}} \equiv \{ (X_{\ell}, Z_{\ell}) \textup{ for } \ell \in [N], \ \mathfrak{M}
\}$, the potential outcomes are mutually independent across all matched units.  
Consequently, for any $\sigma^{i}$s in $\mathcal{S}_2=\{(1,2), (2,1)\}$\footnote{
	For simplicity, here we view $Y^0_{i(\sigma)}$ and $Y^0_{i(\sigma')}$, with $\sigma \ne \sigma'$, as distinct elements when considering the distribution of $Y_{i\bigcdot}$, even though their values can be the same.
	The same logic holds in the later discussion when we consider $Y_{i(j)}^0$s for the distribution of $Y_{i1}^0$ and the $q_{ij}$s for the distribution of $Q_{i1}$.}, 
\begin{align*}%
	& \quad \ \Pr( Y_{1\bigcdot}^0 = Y^0_{1(\sigma^{1})}, \ldots, Y_{I\bigcdot}^0 = Y^0_{I(\sigma^I)} 
	\mid \mathcal{F}_{\text{pair}}
	)
	\nonumber
	\\
	& = 
	\Pr(  Y_{1\bigcdot}^0 = Y^0_{1(\sigma^{1})}, \ldots, Y_{I\bigcdot}^0 = Y^0_{I(\sigma^I)}  
	\mid \mathcal{G}_{\text{pair}}, \ 
	(Y_{i(1)}^0, Y_{i(2)}^0) \text{ for } i \in [I]
	)
	\nonumber
	\\
	& = 
	\prod_{i=1}^I \Pr( Y_{i\bigcdot}^0 = Y^0_{i(\sigma^i)}
	\mid \mathcal{G}_{\text{pair}}, \ (Y_{i(1)}^0, Y_{i(2)}^0)
	)\\
	& = \prod_{i=1}^I \Pr( Y_{i\bigcdot}^0 = Y^0_{i(\sigma^i)}
	\mid \mathcal{F}_{\text{pair}} ). 
	\nonumber
\end{align*}
Thus, conditional on $\mathcal{F}_{\text{pair}}$, the control potential outcomes are mutually independent across all matched pairs. 

Second, we study the conditional distribution of control potential outcomes within each matched pair. 
From the discussion before, we have, for $i \in [I]$, 
\begin{align}\label{eq:r_i_pair_deri}
	\ratio_i & \equiv  
	\frac{
		\Pr( Y_{i\bigcdot}^0 = (Y^0_{i(2)}, Y^0_{i(1)})
		\mid \mathcal{F}_{\text{pair}}
		)
	}{
		\Pr( Y_{i\bigcdot}^0 = (Y^0_{i(1)}, Y^0_{i(2)})
		\mid \mathcal{F}_{\text{pair}}
		)
	}
	= 
	\frac{
		\Pr( Y_{i\bigcdot}^0 = (Y^0_{i(2)}, Y^0_{i(1)})
		\mid \mathcal{G}_{\text{pair}}, \ (Y_{i(1)}^0, Y_{i(2)}^0)
		)
	}{
		\Pr( Y_{i\bigcdot}^0 = (Y^0_{i(1)}, Y^0_{i(2)})
		\mid \mathcal{G}_{\text{pair}}, \ (Y_{i(1)}^0, Y_{i(2)}^0)
		)
	}
	\nonumber
	\\
	& = \frac{p_1( Y^0_{i(2)} \mid X_{i1}) p_0( Y^0_{i(1)} \mid X_{i2})}{
		p_1( Y^0_{i(1)} \mid X_{i1}) p_0( Y^0_{i(2)} \mid X_{i2})
	},
\end{align}
recalling that we assume $Z_{i1}=1$ and $Z_{i2} = 0$ for all $i$, 
and $p_{z}(\cdot \mid x) \equiv p(\cdot \mid x, z)$ denotes the density for the conditional distribution of $Y^\control$ given $X=x$ and $Z=z$, for $z=0,1$. 

Third, we prove the equivalent form of $\ratio_i$s in \eqref{eq:decomp_ps}. 
By Bayes' rule, $p_z(y\mid x) = \pi_z(x, y) p(y\mid x) / \pi_z(x)$, where $\pi_z(x) = \Pr(Z=z\mid X=x)$. 
This then implies that, for any $i \in [I]$, 
\begin{align}\label{eq:r_i_pair_deri_equiv}
	\ratio_i & = \frac{p_1( Y^0_{i(2)} \mid X_{i1}) p_0( Y^0_{i(1)} \mid X_{i2})}{
		p_1( Y^0_{i(1)} \mid X_{i1}) p_0( Y^0_{i(2)} \mid X_{i2})
	}
	\nonumber
	\\
	& = 
	\frac{ 
		\pi_1(X_{i1}, Y^0_{i(2)}) p(Y^0_{i(2)} \mid X_{i1}) / \pi_1(X_{i1}) 
		\cdot
		\pi_0(X_{i2}, Y^0_{i(1)}) p(Y^0_{i(1)}\mid X_{i2}) / \pi_0(X_{i2})
	}{
		\pi_1(X_{i1}, Y^0_{i(1)}) p(Y^0_{i(1)}\mid X_{i1}) / \pi_1(X_{i1})
		\cdot 
		\pi_0(X_{i2}, Y^0_{i(2)}) p(Y^0_{i(2)}\mid X_{i2}) / \pi_0(X_{i2})
	}
	\nonumber
	\\
	& = 
	\frac{ 
		\pi_1(X_{i1}, Y^0_{i(2)}) p(Y^0_{i(2)} \mid X_{i1})
		\cdot
		\pi_0(X_{i2}, Y^0_{i(1)}) p(Y^0_{i(1)}\mid X_{i2})
	}{
		\pi_1(X_{i1}, Y^0_{i(1)}) p(Y^0_{i(1)}\mid X_{i1})
		\cdot 
		\pi_0(X_{i2}, Y^0_{i(2)}) p(Y^0_{i(2)}\mid X_{i2})
	}
	\nonumber
	\\
	& = 
	\frac{ 
		p(Y^0_{i(2)} \mid X_{i1})/p(Y^0_{i(2)}\mid X_{i2})
	}{
		p(Y^0_{i(1)}\mid X_{i1})/p(Y^0_{i(1)}\mid X_{i2})
	}
	\cdot
	\frac{\pi_0(X_{i2}, Y^0_{i(1)})/\pi_0(X_{i2}, Y^0_{i(2)}) }{
		\pi_1(X_{i1}, Y^0_{i(1)})/\pi_1(X_{i1}, Y^0_{i(2)})
	}.
\end{align}
From the above, the equivalent form of $\ratio_i$s in \eqref{eq:decomp_ps} holds. 
Note that, in the above derivation, 
to avoid technical clutter,
we implicitly assume that the terms in the denominators are positive. 

\begin{remark}\label{rmk:iid_weaker_pair}
	The above derivation for the permutation distribution of $Y^0_{i\bigcdot}$s can hold under weaker conditions than Assumption \ref{asmp:iid}. 
	First, the second last equality in \eqref{eq:r_i_pair_deri} still holds when the potential outcomes $Y_\ell^0$s in the original data prior to matching are assumed to be conditionally independent given the covariates $X_\ell$s and treatment assignment $Z_\ell$s. 
	Second, if further, for each unit, the distribution of its control potential outcome depends only on the values of its covariate vector and treatment assignment, 
	then the last equality in \eqref{eq:r_i_pair_deri} will also hold. 
	Third, in addition, if the treatment assignments  $Z_\ell$s are conditionally independent given the covariates $X_\ell$s, and the distribution of $Z_\ell$ depends only on the value of $X_\ell$, 
	then the equivalent forms of $\ratio_i$s in \eqref{eq:r_i_pair_deri_equiv} will still hold. 
	From these, we can know that 
	the conclusions in \S \ref{sec:pair} of the main paper for matched pair studies can hold under weaker conditions than Assumption \ref{asmp:iid}. 
	For example, we can weaken Assumption \ref{asmp:iid} by allowing fixed, instead of i.i.d., covariates.
\end{remark}

\subsection{Proof of Theorem \ref{thm:exact_pair}}

Consider any test statistic $t(Z_{\bigcdot\bigcdot}, Y^0_{\bigcdot\bigcdot}) = \sum_{i=1}^I \sum_{j=1}^{2} Z_{ij} Q_{ij}$ that satisfies Condition \ref{cond:symmetry}. 
Recall that we assume $Z_{i1}=1$ and $Z_{i2} = 0$ for all $i$. 
Given $\mathcal{F}_{\text{pair}}$, 
$t(Z_{\bigcdot\bigcdot}, Y^0_{\bigcdot\bigcdot})$ 
becomes 
$\sum_{i=1}^I Q_{i1}$, and moreover, for all $i$, 
$Q_{i1}$ is $\phi_{i}( 
Y^0_{i1}, 
(Y^0_{1(1)}, Y^0_{1(2)}), 
\ldots, 
(Y^0_{I(1)}, Y^0_{I(2)})
)$. 

First, from \S \ref{sec:tech_detail_pair}, 
the $Y_{i1}$s are mutually independent across all matched pairs. 
This implies that, conditional on $\mathcal{F}_{\text{pair}}$, 
the $Q_{i1}$s becomes mutually independent. 

Second, conditional on $\mathcal{F}_{\text{pair}}$, for each $i$, $Q_{i1}$ takes value in $\{q_{i1}, q_{i2}\}$. 
Moreover, 
\begin{align*}
	\frac{\Pr(Q_{i1} = q_{i2} \mid \mathcal{F}_{\text{pair}})}{
		\Pr(Q_{i1} = q_{i1} \mid \mathcal{F}_{\text{pair}})
	}
	= 
	\frac{\Pr(Y^0_{i1} = Y^0_{i(2)} \mid \mathcal{F}_{\text{pair}})}{
		\Pr(Y^0_{i1} = Y^0_{i(1)} \mid \mathcal{F}_{\text{pair}})
	}
	= \ratio_i. 
\end{align*}
This implies that $\Pr(Q_{i1} = q_{i1} \mid \mathcal{F}_{\text{pair}}) = 1/(1+\ratio_i)$ and 
$\Pr(Q_{i1} = q_{i2} \mid \mathcal{F}_{\text{pair}}) = \ratio_i/(1+\ratio_i)$.

From the above, Theorem \ref{thm:exact_pair} holds.

\subsection{Proof of Theorem \ref{thm:link_perm_rand}}\label{sec:proof_link_perm_rand}

First, under the hypothetical experiment with treatment assignment mechanism described in \S \ref{sec:link_pair},  
with $q_{ij}$s treated as constants, 
$\sum_{j=1}^2 A_{ij} q_{ij}$s are mutually independent across all $i$. 
Moreover, for each $i$, 
$\sum_{j=1}^2 A_{ij} q_{ij}$ takes value $q_{i1}$ with probability $\Pr(A_{i1} = 1) = 1/(1+\ratio_i)$ and $q_{i2}$ with probability $1-1/(1+\ratio_i) = \ratio_i / (1+\ratio_i)$. 
From Theorem \ref{thm:exact_pair}, the conditional distribution of 
$t(Z_{\bigcdot\bigcdot}, Y^0_{\bigcdot\bigcdot})$ given $\mathcal{F}_{\text{pair}}$ is the same as the randomization distribution of 
$t(A_{\bigcdot\bigcdot}, Y^0_{\bigcdot(\bigcdot)}) = \sum_{i=1}^I \sum_{j=1}^2 A_{ij} q_{ij}$, 
where $Y^0_{\bigcdot(\bigcdot)}$ and consequently $q_{ij}$s are treated as constants. 

Second, we show that the probability mass function of $A_{\bigcdot\bigcdot}$ described in \S \ref{sec:link_pair} has the form in \eqref{eq:dist_A} for some $u_{ij}$s in $[0, 1]$. 
For each $i$, if $\ratio_i \ge 1$, define $u_{i1} = 0$ and $u_{i2} = 1$;  
otherwise, define $u_{i1}=1$ and $u_{i2} = 0$. 
We then have, for each $i$, 
\begin{align*}
	\frac{\exp(\gamma_i u_{i2})}{\exp(\gamma_i u_{i1})} 
	= 
	\Gamma_i^{u_{i2}-u_{i1}}
	= 
	\begin{cases}
		\Gamma_i = \ratio_i,  & \text{if } \ratio_i \ge 1, \\
		\Gamma_i^{-1} = (1/\ratio_i)^{-1} = \ratio_i, & \text{if } \ratio_i < 1. 
	\end{cases}
\end{align*}
Consequently, the probability mass function of $A_{\bigcdot\bigcdot}$ described in \S \ref{sec:link_pair} has the following equivalent forms:
\begin{align}\label{eq:pmf_A_pair}
	\Pr(A_{\bigcdot\bigcdot} = a_{\bigcdot\bigcdot})
	& = 
	\prod_{i=1}^I \left(\frac{1}{1+\ratio_i}\right)^{a_{i1}} \left(\frac{\ratio_i}{1+\ratio_i}\right)^{a_{i2}}
	\\
	& = 
	\prod_{i=1}^I  \left\{ \frac{1}{1+\exp(\gamma_i u_{i2})/\exp(\gamma_i u_{i1})} \right\}^{a_{i1}} \left\{ \frac{\exp(\gamma_i u_{i2})/\exp(\gamma_i u_{i1})}{1+\exp(\gamma_i u_{i2})/\exp(\gamma_i u_{i1})}\right\}^{a_{i2}}
	\nonumber
	\\
	& = 
	\prod_{i=1}^I  \left\{ \frac{\exp(\gamma_i u_{i1})}{\exp(\gamma_i u_{i1})+\exp(\gamma_i u_{i2})}\right\}^{a_{i1}} \left\{\frac{\exp(\gamma_i u_{i2})}{\exp(\gamma_i u_{i1})+\exp(\gamma_i u_{i2})}\right\}^{a_{i2}}
	\nonumber
	\\
	& = 
	\prod_{i=1}^I  \frac{\exp(a_{i1}\gamma_i u_{i1})\cdot\exp(a_{i2}\gamma_i u_{i2})}{\exp(\gamma_i u_{i1})+\exp(\gamma_i u_{i2})}
	= 
	\prod_{i=1}^I  \frac{\exp(\gamma_i\sum_{j=1}^2a_{ij} u_{ij})}{\sum_{j=1}^2 \exp(\gamma_i u_{ij})},
	\nonumber
\end{align}
if $a_{ij}$s are all in $\{0,1\}$ and $a_{i1}+a_{i2}=1$ for all $i$, and 
$\Pr(A_{\bigcdot\bigcdot} = a_{\bigcdot\bigcdot}) = 0$ otherwise. 
Thus, the probability mass function of $A_{\bigcdot\bigcdot}$ has the form in \eqref{eq:dist_A} for some $u_{ij}$s in $[0, 1]$. 

From the above, Theorem \ref{thm:link_perm_rand} holds.

\subsection{Proof of Theorem \ref{thm:link_perm_rand_adaptive}}

From the proof of Theorem \ref{thm:link_perm_rand} in \S \ref{sec:proof_link_perm_rand}, 
we know that the conditional distribution of 
$t(Z_{\bigcdot\bigcdot}, Y^0_{\bigcdot\bigcdot})$ given $\mathcal{F}_{\text{pair}}$ is the same as the randomization distribution of 
$t(A_{\bigcdot\bigcdot}, Y^0_{\bigcdot(\bigcdot)}) = \sum_{i=1}^I \sum_{j=1}^2 A_{ij} q_{ij}$, 
where $Y^0_{\bigcdot(\bigcdot)}$ and consequently $q_{ij}$s are treated as constants, 
and $A_{\bigcdot\bigcdot}$ follows the distribution in \eqref{eq:pmf_A_pair}. 
By the definition of $\hat{R}_{\text{m}i}$s and $\tilde{R}_{\text{u}i}$s, 
the probability mass function of $A_{\bigcdot\bigcdot}$ in \eqref{eq:pmf_A_pair} has the following equivalent forms:
\begin{align*}
	\Pr(A_{\bigcdot\bigcdot} = a_{\bigcdot\bigcdot})
	& = 
	\prod_{i=1}^I \left(\frac{1}{1+\ratio_i}\right)^{a_{i1}} \left(\frac{\ratio_i}{1+\ratio_i}\right)^{a_{i2}}
	= 
	\prod_{i=1}^I \left(\frac{1}{1+\hat{\ratio}_{\text{m}i} \tilde{\ratio}_{\text{u}i} }\right)^{a_{i1}} \left(\frac{\hat{\ratio}_{\text{m}i} \tilde{\ratio}_{\text{u}i}}{1+\hat{\ratio}_{\text{m}i} \tilde{\ratio}_{\text{u}i}}\right)^{a_{i2}},
\end{align*}
if $a_{ij}$s are all in $\{0,1\}$ and $a_{i1}+a_{i2}=1$ for all $i$, and 
$\Pr(A_{\bigcdot\bigcdot} = a_{\bigcdot\bigcdot}) = 0$ otherwise. 
For each $i$, if $\tilde{\ratio}_{\text{u}i} \ge 1$, define $u_{i1} = 0$ and $u_{i2} = 1$; 
otherwise, define $u_{i1}=1$ and $u_{i2} = 0$. 
Then, by the definition of $\tilde{\Gamma}_{\text{u}i}$s, we have 
\begin{align*}
	\frac{\exp(\tilde{\gamma}_{\text{u}i} u_{i2})}{\exp(\tilde{\gamma}_{\text{u}i} u_{i1})} 
	= 
	\tilde{\Gamma}_{\text{u}i}^{u_{i2}-u_{i1}}
	= 
	\begin{cases}
		\tilde{\Gamma}_{\text{u}i} = \tilde{\ratio}_{\text{u}i},  & \text{if } \tilde{\ratio}_{\text{u}i} \ge 1, \\
		\tilde{\Gamma}_{\text{u}i}^{-1} = (1/\tilde{\ratio}_{\text{u}i})^{-1} = \tilde{\ratio}_{\text{u}i}, & \text{if } \tilde{\ratio}_{\text{u}i} < 1. 
	\end{cases}
\end{align*}
We then have the following equivalent forms for the probability mass function of $A_{\bigcdot\bigcdot}$: 
\begin{align*}
	& \quad \ \Pr(A_{\bigcdot\bigcdot} = a_{\bigcdot\bigcdot})
	\\
	& 
	= 
	\prod_{i=1}^I \left(\frac{1}{1+\hat{\ratio}_{\text{m}i} \tilde{\ratio}_{\text{u}i} }\right)^{a_{i1}} \left(\frac{\hat{\ratio}_{\text{m}i} \tilde{\ratio}_{\text{u}i}}{1+\hat{\ratio}_{\text{m}i} \tilde{\ratio}_{\text{u}i}}\right)^{a_{i2}}
	\\
	& = 
	\prod_{i=1}^I \left(\frac{1}{1+\hat{\ratio}_{\text{m}i} \exp(\tilde{\gamma}_{\text{u}i} u_{i2})/\exp(\tilde{\gamma}_{\text{u}i} u_{i1}) }\right)^{a_{i1}} \left(\frac{\hat{\ratio}_{\text{m}i} \exp(\tilde{\gamma}_{\text{u}i} u_{i2})/\exp(\tilde{\gamma}_{\text{u}i} u_{i1})}{1+\hat{\ratio}_{\text{m}i} \exp(\tilde{\gamma}_{\text{u}i} u_{i2})/\exp(\tilde{\gamma}_{\text{u}i} u_{i1})}\right)^{a_{i2}}\\
	& = 
	\prod_{i=1}^I \left(
	\frac{\exp(\tilde{\gamma}_{\text{u}i} u_{i1})}{\exp(\tilde{\gamma}_{\text{u}i} u_{i1}) + \exp(\log \hat{\ratio}_{\text{m}i} + \tilde{\gamma}_{\text{u}i} u_{i2}) }
	\right)^{a_{i1}} \left(\frac{ \exp(\log \hat{\ratio}_{\text{m}i} + \tilde{\gamma}_{\text{u}i} u_{i2})}{\exp(\tilde{\gamma}_{\text{u}i} u_{i1})+ \exp(\log \hat{\ratio}_{\text{m}i} + \tilde{\gamma}_{\text{u}i} u_{i2})}\right)^{a_{i2}}\\
	& = 
	\prod_{i=1}^I 
	\frac{
		\exp\{ a_{i1}\tilde{\gamma}_{\text{u}i} u_{i1}
		+ 
		a_{i2} (\log \hat{\ratio}_{\text{m}i} + \tilde{\gamma}_{\text{u}i} u_{i2})
		\}
	}{
		\exp(\tilde{\gamma}_{\text{u}i} u_{i1})+ \exp(\log \hat{\ratio}_{\text{m}i} + \tilde{\gamma}_{\text{u}i} u_{i2})
	},
\end{align*}
if $a_{ij}$s are all in $\{0,1\}$ and $a_{i1}+a_{i2}=1$ for all $i$, and 
$\Pr(A_{\bigcdot\bigcdot} = a_{\bigcdot\bigcdot}) = 0$ otherwise. 

From the above, we can then derive Theorem \ref{thm:link_perm_rand_adaptive}.

\subsection{Adaptive inference for matched pair studies}\label{sec:supp_adap_pair}

We consider sensitivity analysis with a uniform bound on the $\tilde{\Gamma}_{\text{u}i}$s for all matched pairs. 
To achieve valid tests for the sharp null $H_0$, 
we need to derive the worst-case permutation distribution of $t(Z_{\bigcdot\bigcdot}, Y^0_{\bigcdot\bigcdot})$. 
From Theorem \ref{thm:link_perm_rand_adaptive}, 
it suffices to consider the worst-case randomization distribution of $t(A_{\bigcdot\bigcdot}, Y^0_{\bigcdot(\bigcdot)}) = \sum_{i=1}^I \sum_{j=1}^2 A_{ij} q_{ij}$ under the constraint that $\tilde{\Gamma}_{\text{u}i}\le \Lambda$ for all $i$ and some prespecified $\Lambda \ge 1$, 
where $A_{\bigcdot\bigcdot}$ follows the distribution in \eqref{eq:cov_adap_pair} for some $u_{ij}$s in $[0,1]$. 
For each $i\in [I]$, we know that $\sum_{j=1}^2 A_{ij} q_{ij}$ takes value in $\{q_{i1}, q_{i2}\}$, and it equals $q_{i2}$ with probability 
\begin{align*}
	\Pr(A_{i2} = 1) & = \frac{ \exp(\log \hat{\ratio}_{\text{m}i} + \tilde{\gamma}_{\text{u}i} u_{i2})}{\exp(\tilde{\gamma}_{\text{u}i} u_{i1})+ \exp(\log \hat{\ratio}_{\text{m}i} + \tilde{\gamma}_{\text{u}i} u_{i2})}
	\begin{cases}
		\ge \frac{ \exp(\log \hat{\ratio}_{\text{m}i} )}{\exp(\tilde{\gamma}_{\text{u}i})+ \exp(\log \hat{\ratio}_{\text{m}i})}
		=  \frac{\hat{\ratio}_{\text{m}i}}{\tilde{\Gamma}_{\text{u}i} + \hat{\ratio}_{\text{m}i}}
		\ge \frac{\hat{\ratio}_{\text{m}i}}{\Lambda + \hat{\ratio}_{\text{m}i}}
		, 
		\\
		\le 
		\frac{ \exp(\log \hat{\ratio}_{\text{m}i} + \tilde{\gamma}_{\text{u}i})}{\exp(0)+ \exp(\log \hat{\ratio}_{\text{m}i} + \tilde{\gamma}_{\text{u}i})}
		= 
		\frac{  \hat{\ratio}_{\text{m}i} \tilde{\Gamma}_{\text{u}i}
		}{1 + \hat{\ratio}_{\text{m}i} \tilde{\Gamma}_{\text{u}i}}
		\le \frac{  \hat{\ratio}_{\text{m}i} \Lambda
		}{1 + \hat{\ratio}_{\text{m}i} \Lambda}
		, 
	\end{cases}    
\end{align*}
where the bounds are achieved when $\tilde{\Gamma}_{\text{u}i} = \Lambda$ and $(u_{i1}, u_{i2})$ equals $(1,0)$ or $(0,1)$. 
For each $i$, 
let $\overline{A}_{i2}$ be a Bernoulli random variable with 
\begin{align*}
	\Pr(\overline{A}_{i2} = 1) = 
	\begin{cases}
		\frac{\hat{\ratio}_{\text{m}i}}{\Lambda + \hat{\ratio}_{\text{m}i}}, & \text{if } q_{i2} < q_{i1}, \\
		\frac{  \hat{\ratio}_{\text{m}i} \Lambda
		}{1 + \hat{\ratio}_{\text{m}i} \Lambda}, 
		& \text{if } q_{i2} \ge q_{i1}, 
	\end{cases}
\end{align*}
and $\overline{A}_{i1} = 1-\overline{A}_{i2}$. 
It is not difficult to see that 
$\sum_{j=1}^2 A_{ij} q_{ij}$ is stochastically smaller than or equal to $\sum_{j=1}^2 \overline{A}_{ij} q_{ij}$. 
We further let $(\overline{A}_{i1}, \overline{A}_{i2})$ be mutually independent across all $i$. 
From \citet[][Lemma A3]{WL23}, 
$\sum_{i=1}^I \sum_{j=1}^2 A_{ij} q_{ij}$ is stochastically smaller than or equal to $\sum_{i=1}^I \sum_{j=1}^2 \overline{A}_{ij} q_{ij}$. 
Therefore, the distribution of $\sum_{i=1}^I \sum_{j=1}^2 \overline{A}_{ij} q_{ij}$ is the worst-case permutation distribution of the test statistic under the constraint that $\tilde{\Gamma}_{\text{u}i}\le \Lambda$ for all $i$. 
A valid p-value for $H_0$ can then be obtained by evaluating the tail probability of $\sum_{i=1}^I \sum_{j=1}^2 \overline{A}_{ij} q_{ij}$ at the realized value of the test statistic.

	As a side note, the sensitivity analysis described in this subsection contains that in \S \ref{sec:link_pair} as a special case, where we can simply set all the $\hat{R}_{\text{m}i}$s to be $1$.

\section{Technical details for matched set studies}

\subsection{Technical details for \S \ref{sec:set}}\label{sec:tech_set}

Below we study the conditional distribution of the potential outcomes $Y^0_{i1}$s given $\mathcal{F}_{\text{set}}$, recalling that we assume $Z_{i1} = 1$ and $Z_{i2} = \ldots = Z_{i n_i} = 0$ for all matched sets $i$.

First, by the same logic as that in \S \ref{sec:tech_detail_pair}, 
conditional on $\mathcal{F}_{\text{set}}$, 
the control potential outcomes $Y_{i\bigcdot}^0$s are mutually independent across all matched sets. 
Moreover, for each set $i$, 
the conditional distribution of $Y_{i\bigcdot}^0$ given $\mathcal{F}_{\text{set}}$ satisfies that, for any $\sigma \in \mathcal{S}_{n_i}$, recalling that $\mathcal{S}_{n_i}$ is the set of all permutations of $(1,2,\ldots, n_i)$, 
\begin{align*}
	\Pr(Y_{i\bigcdot}^0 = Y^0_{i(\sigma)}\mid \mathcal{F}_{\text{set}})
	\propto 
	p_1(Y^0_{i(\sigma_1)}\mid X_{i1}) \prod_{m=2}^{n_i} p_0( Y^0_{i(\sigma_m)} \mid X_{im}).
\end{align*}
By Bayes' rule, $p_z(y\mid x) = \pi_z(x, y) p(y\mid x) / \pi_z(x)$, where $\pi_z(x) = \Pr(Z=z\mid X=x)$. 
We then have, for any set $i$ and any $\sigma \in \mathcal{S}_{n_i}$, 
\begin{align*}
	\Pr(Y_{i\bigcdot}^0 = Y^0_{i(\sigma)}\mid \mathcal{F}_{\text{set}})
	& \propto 
	\frac{\pi_1(X_{i1}, Y^0_{i(\sigma_1)}) p(Y^0_{i(\sigma_1)}\mid X_{i1})}{\pi_1(X_{i1})}
	\cdot
	\prod_{m=2}^{n_i} 
	\frac{\pi_0(X_{im}, Y^0_{i(\sigma_m)} ) p(Y^0_{i(\sigma_m)} \mid X_{im}) }{\pi_0(X_{im})}\\
	& \propto 
	\underbrace{ \prod_{m=1}^{n_i}  p(Y^0_{i(\sigma_m)} \mid X_{im})}_{p_{i\sigma}}
	\cdot
	\underbrace{ \pi_1(X_{i1}, Y^0_{i(\sigma_1)}) \prod_{m=2}^{n_i} \pi_0(X_{im}, Y^0_{i(\sigma_m)} ) }_{\pi_{i\sigma}} 
	= p_{i\sigma} \cdot \pi_{i\sigma},
\end{align*}
where $p_{i\sigma}$ and $\pi_{i\sigma}$ are defined as in \S \ref{sec:bias_set}.

Second, for each $i$, conditional $\mathcal{F}_{\text{set}}$, 
$Y^0_{i1}$ takes value in $\{Y^0_{i(1)}, \ldots, Y^0_{i(n_i)}\}$. 
Below we consider the ratio between 
$\Pr(Y^0_{i1} = Y^0_{i(j)} \mid \mathcal{F}_{\text{set}})$
and 
$\Pr(Y_{i1} = Y^0_{i(k)} \mid \mathcal{F}_{\text{set}})$
for any $1\le j\ne k \le n_i$. 
From the discussion before, we know that 
\begin{align}\label{eq:ratio_set_supp}
	\frac{\Pr(Y^0_{i1} = Y^0_{i(j)} \mid \mathcal{F}_{\text{set}})}{
		\Pr(Y^0_{i1} = Y^0_{i(k)} \mid \mathcal{F}_{\text{set}})
	}
	= 
	\frac{
		\sum_{\sigma \in \mathcal{S}_{n_i}: \sigma_1 = j}
		\Pr( Y_{i\bigcdot}^0 = Y^0_{i(\sigma)} \mid \mathcal{F}_{\textup{set}} )
	}{
		\sum_{\sigma \in \mathcal{S}_{n_i}: \sigma_1 = k}
		\Pr( Y_{i\bigcdot}^0 = Y^0_{i(\sigma)} \mid \mathcal{F}_{\textup{set}} )
	}
	=
	\frac{
		\sum_{\sigma \in \mathcal{S}_{n_i}: \sigma_1 = j}
		p_{i\sigma} \pi_{i\sigma}
	}{
		\sum_{\sigma \in \mathcal{S}_{n_i}: \sigma_1 = k}
		p_{i\sigma} \pi_{i\sigma}
	}
\end{align}

Third, we give a bound for \eqref{eq:ratio_set_supp} that aligns more
closely with \citet{Rosenbaum02a}’s measure of unmeasured confounding strength. 
For any permutation $\sigma\in \mathcal{S}_{n_i}$ with $\sigma_1 = j$, 
we define $\sigma'$ the same as $\sigma$ except that it switches the positions of $j$ and $k$. 
Specifically, if $\sigma_1=j$ and $\sigma_l = k$, 
then $\sigma'_1 =k$,  $\sigma'_l = j$, and $\sigma'_m = \sigma_m$ for $m \in [n_i]\setminus\{1,l\}$. 
We can verify that (i) $\sigma' \in \mathcal{S}_{n_i}$ and $\sigma_1' = k$, 
and (ii) when $\sigma$ traverses all permutations in $\mathcal{S}_{n_i}$ with $j$ as their first elements, 
the corresponding $\sigma'$ traverses all permutations in $\mathcal{S}_{n_i}$ with $k$ as their first elements. 
Consequently, we can bound the ratio between $\Pr(Y^0_{i1} = Y^0_{i(j)} \mid \mathcal{F}_{\text{set}})$
and 
$\Pr(Y^0_{i1} = Y^0_{i(k)} \mid \mathcal{F}_{\text{set}})$
by 
\begin{align}\label{eq:bound_align_Rosen}
	\frac{\Pr(Y_{i1}^0 = Y^0_{i(j)}\mid \mathcal{F}_{\textup{set}})}{
		\Pr(Y_{i1}^0 = Y^0_{i(k)}\mid \mathcal{F}_{\textup{set}})
	}
	& = 
	\frac{
		\sum_{\sigma \in \mathcal{S}_{n_i}: \sigma_1 = j}
		p_{i\sigma} \pi_{i\sigma}
	}{
		\sum_{\sigma \in \mathcal{S}_{n_i}: \sigma_1 = k}
		p_{i\sigma} \pi_{i\sigma}
	}
	= 
	\frac{
		\sum_{\sigma \in \mathcal{S}_{n_i}: \sigma_1 = j}
		p_{i\sigma} \pi_{i\sigma}
	}{
		\sum_{\sigma \in \mathcal{S}_{n_i}: \sigma_1 = j}
		p_{i\sigma'} \pi_{i\sigma'}
	}
	\le 
	\max_{\sigma \in \mathcal{S}_{n_i}: \sigma_1 = j}
	\frac{
		p_{i\sigma} \pi_{i\sigma}
	}{
		p_{i\sigma'} \pi_{i\sigma'}
	}.
\end{align}
For any $\sigma \in \mathcal{S}_{n_i}$ with $\sigma_1 = j$, 
let $l$ denote the index in $\{2, \ldots, n_i\}$ such that $\sigma_l = k$. 
By the definition of $\sigma'$, we have $\sigma_m = \sigma'_m$ for $m\notin \{1, l\}$, 
$\sigma'_1 = k$ and $\sigma'_l = j$. 
Thus, 
\begin{align}\label{eq:ratio_p_pi_sigma}
	\frac{
		p_{i\sigma} \pi_{i\sigma}
	}{
		p_{i\sigma'} \pi_{i\sigma'}
	}
	& = 
	\frac{
		\prod_{m=1}^{n_i}  p(Y^0_{i(\sigma_m)} \mid X_{im})
		\cdot
		\pi_1(X_{i1}, Y^0_{i(\sigma_1)}) \prod_{m=2}^{n_i} \pi_0(X_{im}, Y^0_{i(\sigma_m)} )
	}{
		\prod_{m=1}^{n_i}  p(Y^0_{i(\sigma'_m)} \mid X_{im})
		\cdot
		\pi_1(X_{i1}, Y^0_{i(\sigma'_1)}) \prod_{m=2}^{n_i} \pi_0(X_{im}, Y^0_{i(\sigma'_m)} )    
	}
	\nonumber
	\\
	& = 
	\frac{
		p(Y^0_{i(j)} \mid X_{i1})p(Y^0_{i(k)} \mid X_{il})
		\cdot
		\pi_1(X_{i1}, Y^0_{i(j)}) \pi_0(X_{il}, Y^0_{i(k)} )
	}{
		p(Y^0_{i(k)} \mid X_{i1})p(Y^0_{i(j)} \mid X_{il})
		\cdot
		\pi_1(X_{i1}, Y^0_{i(k)}) \pi_0(X_{il}, Y^0_{i(j)} ) 
	}
	\nonumber
	\\
	& = 
	\frac{
		p(Y^0_{i(j)} \mid X_{i1})/p(Y^0_{i(j)} \mid X_{il})
		\cdot
		\pi_0(X_{il}, Y^0_{i(k)} )/\pi_0(X_{il}, Y^0_{i(j)} ) 
	}{
		p(Y^0_{i(k)} \mid X_{i1})/p(Y^0_{i(k)} \mid X_{il})
		\cdot
		\pi_1(X_{i1}, Y^0_{i(k)}) / \pi_1(X_{i1}, Y^0_{i(j)})
	}
	\equiv \ratio_{ijkl}.
\end{align}
This immediately implies that, 
for any $1 \le j\ne k \le n_i$, 
$\Pr(Y_{i1}^0 = Y^0_{i(j)}\mid \mathcal{F}_{\textup{set}})/\Pr(Y_{i1}^0 = Y^0_{i(k)}\mid \mathcal{F}_{\textup{set}}) \le \max_{2\le l \le n_i} \ratio_{ijkl}$. 
Define $\bar{\Gamma}_i = \max_{1\le j \ne k \le n_i, 2\le l \le n_i} \ratio_{ijkl}$. Because $\ratio_{ijkl} = 1/\ratio_{ikjl}$ by definition, we obviously have $\bar{\Gamma}_i\ge 1$ for all $i$. 
From the above, we further have, for any $i\in [I]$ and $j,k\in [n_i]$, 
\begin{align*}
	\bar{\Gamma}_i^{-1} \le 
	\frac{\Pr(Y_{i1}^0 = Y^0_{i(j)}\mid \mathcal{F}_{\textup{set}})}{
		\Pr(Y_{i1}^0 = Y^0_{i(k)}\mid \mathcal{F}_{\textup{set}})
	}
	& \le \bar{\Gamma}_i. 
\end{align*}
The $\bar{\Gamma}_i$s here play a similar role as $\Gamma_i$s in \eqref{eq:set_prob_ratio}, and they quantify the biases from uniform permutation. 
Moreover, the expression of $\bar{\Gamma}_i$s aligns more closely with the bias measure in \citet{Rosenbaum02a}.
As discussed in \S \ref{sec:simp_special}, 
in the case of exact matching, 
we can further simplify the $\ratio_{ijkl}$s and the corresponding $\bar{\Gamma}_i$s, which essentially reduce to Rosenbaum's measure on the strength of unmeasured confounding.

Fourth, we give the technical details for \eqref{eq:match_quality_set}--\eqref{eq:set_prob_ratio}. 
By definition and applying the same idea as in \eqref{eq:bound_align_Rosen} and \eqref{eq:ratio_p_pi_sigma}, we have, for all $i\in [I]$, 
\begin{align*}
	\Gamma_{\text{m}i} 
	& \equiv
	\max_{j,k\in [n_i]} 
	\frac{p_{\text{m}ij}}{p_{\text{m}ik}}
	= 
	\max_{1\le j\ne k\in n_i} \frac{
		\sum_{\sigma \in \mathcal{S}_{n_i}: \sigma_1 = j}
		p_{i\sigma}
	}{
		\sum_{\sigma \in \mathcal{S}_{n_i}: \sigma_1 = k}
		p_{i\sigma}
	}
	\le 
	\max_{1\le j\ne k \le n_i} 
	\max_{\sigma \in \mathcal{S}_{n_i}: \sigma_1 = j}
	\frac{
		p_{i\sigma} 
	}{
		p_{i\sigma'}
	}\\
	& 
	\le 
	\max_{1\le j\ne k \le n_i} 
	\max_{2 \le l \le n_i}
	\frac{
		p(Y^0_{i(j)} \mid X_{i1})/p(Y^0_{i(j)} \mid X_{il})
	}{
		p(Y^0_{i(k)} \mid X_{i1})/p(Y^0_{i(k)} \mid X_{il})
	}
	= \max_{j,k,l\in [n_i]} 
	\frac{
		p(Y^0_{i(j)} \mid X_{i1})/p(Y^0_{i(j)} \mid X_{il})
	}{
		p(Y^0_{i(k)} \mid X_{i1})/p(Y^0_{i(k)} \mid X_{il})
	}, 
\end{align*}
where $\sigma'$ is the same as $\sigma$ except that it switches the positions of $j$ and $k$.
This immediately implies  \eqref{eq:match_quality_set} in the main paper. 
By definition, for each $i\in [I]$, 
\begin{align*}
	\Gamma_{\text{u}i} & \equiv \frac{
		\max_{\sigma \in \mathcal{S}_{n_i}} \pi_{i\sigma} 
	}{
		\min_{\psi \in \mathcal{S}_{n_i}} \pi_{i\psi} 
	} 
	= 
	\frac{
		\max_{\sigma \in \mathcal{S}_{n_i}} 
		\pi_1(X_{i1}, Y^0_{i(\sigma_1)}) \prod_{m=2}^{n_i} \pi_0(X_{im}, Y^0_{i(\sigma_m)} )
	}{
		\min_{\psi \in \mathcal{S}_{n_i}} 
		\pi_1(X_{i1}, Y^0_{i(\psi_1)}) \prod_{m=2}^{n_i} \pi_0(X_{im}, Y^0_{i(\psi_m)} )
	} \\
	& \le 
	\max_{j,k\in [n_i]}
	\frac{\pi_1(X_{i1}, Y^0_{i(j)})}{\pi_1(X_{i1}, Y^0_{i(k)})} 
	\prod_{m=2}^{n_i}
	\left[ 
	\max_{j,k\in [n_i]}
	\frac{\pi_0(X_{im}, Y^0_{i(j)})}{
		\pi_0(X_{im}, Y^0_{i(k)} )
	}
	\right],
\end{align*}
which immediately implies \eqref{eq:confound_strength_set} in the main paper. 
From \eqref{eq:bound_align_Rosen}, for any $i\in [I]$ and $j,k\in [n_i]$, 
\begin{align*}
	\frac{\Pr(Y_{i1}^0 = Y^0_{i(j)}\mid \mathcal{F}_{\textup{set}})}{
		\Pr(Y_{i1}^0 = Y^0_{i(k)}\mid \mathcal{F}_{\textup{set}})
	}
	& = 
	\frac{
		\sum_{\sigma \in \mathcal{S}_{n_i}: \sigma_1 = j}
		p_{i\sigma} \pi_{i\sigma}
	}{
		\sum_{\psi \in \mathcal{S}_{n_i}: \psi_1 = k}
		p_{i\psi} \pi_{i\psi}
	}
	= 
	\frac{p_{\text{m}ij}}{p_{\text{m}ik}} \cdot 
	\frac{
		\sum_{\sigma \in \mathcal{S}_{n_i}: \sigma_1 = j}
		p_{i\sigma}/p_{\text{m}ij} \cdot \pi_{i\sigma}
	}{
		\sum_{\psi \in \mathcal{S}_{n_i}: \psi_1 = k}
		p_{i\psi}/p_{\text{m}ik} \cdot \pi_{i\psi}
	}
	\\
	& \in 
	\frac{p_{\text{m}ij}}{p_{\text{m}ik}} \cdot 
	\left[
	\frac{
		\min_{\psi \in \mathcal{S}_{n_i}} \pi_{i\psi} 
	}{
		\max_{\sigma \in \mathcal{S}_{n_i}} \pi_{i\sigma} 
	}, \ \ 
	\frac{
		\max_{\sigma \in \mathcal{S}_{n_i}} \pi_{i\sigma} 
	}{
		\min_{\psi \in \mathcal{S}_{n_i}} \pi_{i\psi} 
	} 
	\right]
	= \frac{p_{\text{m}ij}}{p_{\text{m}ik}} \cdot \big[ 
	\Gamma_{\text{u}i}^{-1}, \  
	\Gamma_{\text{u}i}
	\big]\\
	& \in 
	\big[ 
	\Gamma_{\text{m}i}^{-1} \Gamma_{\text{u}i}^{-1}, \  
	\Gamma_{\text{m}i}  \Gamma_{\text{u}i}
	\big]
	= \big[ 
	\Gamma_i^{-1}, \  
	\Gamma_i
	\big], 
\end{align*}
where the last two steps follow by the definition of $\Gamma_{\text{m}i}$ and $\Gamma_{i}$. 
This immediately implies \eqref{eq:set_prob_ratio} in the main paper.

Note that, similar to \S \ref{sec:tech_detail_pair}, to avoid technical clutter, 
we implicitly assume that the terms in the denominators in the above derivation are positive. 

\begin{remark}
	Similar to Remark \ref{rmk:iid_weaker_pair}, the above derivation can hold under weaker conditions than Assumption \ref{asmp:iid}. 
	Specifically, we can relax Assumption \ref{asmp:iid} for matched set studies in the same way as in Remark \ref{rmk:iid_weaker_pair} for matched pair studies. 
\end{remark}

\subsection{Proof of Theorem \ref{thm:perm_set}}
Consider any test statistic $t(Z_{\bigcdot \bigcdot}, Y_{\bigcdot \bigcdot}^0) = \sum_{i=1}^I \sum_{j=1}^{n_i} Z_{ij} Q_{ij}$ that satisfies Condition \ref{cond:symmetry_set}.
Recall that we assume, without loss of generality, that $Z_{i1} =1$ and $Z_{ij} = 0$ for $2\le j\le n_i$. 
Conditional on $\mathcal{F}_{\text{set}}$, 
$t(Z_{\bigcdot \bigcdot}, Y_{\bigcdot \bigcdot}^0)$
becomes 
$\sum_{i=1}^I Q_{i1}$, and moreover, 
$Q_{i1}$ is $\phi_i(Y_{i1}^0, Y^0_{1(\bigcdot)}, \ldots, Y^0_{I(\bigcdot)})$.

From \S \ref{sec:tech_set}, conditional on $\mathcal{F}_{\text{set}}$, 
$Q_{i1}$s are mutually independent across all $i \in [I]$. 
Furthermore, for each $i$, $Q_{i1}$ takes value in $\{q_{i1}, \ldots, q_{in_i}\}$, 
and 
$Q_{i1}$ equals $q_{ij}$ with probability $\Pr(Y_{i1}^0 = Y^0_{i(j)} \mid \mathcal{F}_{\text{set}})$. 
This immediately implies that, for all $i$ and $1\le j\ne k \le n_i$, 
$$
\Gamma_i^{-1} \le \Pr(Q_{i1} = q_{ij}\mid \mathcal{F}_{\text{set}})/\Pr(Q_{i1} = q_{ik}\mid \mathcal{F}_{\text{set}}) \le \Gamma_i.
$$

From the above, Theorem \ref{thm:perm_set} holds.

\subsection{Proof of Theorem \ref{thm:link_perm_rand_set}}

For any $i \in [I]$, 
define 
\begin{align}\label{eq:u_ij_construct_set}
	u_{ij} = \frac{1}{\gamma_i}\log\left\{ \frac{\Pr(Y^0_{i1} = Y^0_{i(j)}\mid \mathcal{F}_{\text{set}})}{\min_{1\le k \le n_i}\Pr(Y^0_{i1} = Y^0_{i(k)}\mid \mathcal{F}_{\text{set}})} \right\} 
	\qquad (j \in  [n_i])
\end{align}
if $\gamma_i = \log \Gamma_i > 0$, and $u_{i1} = \ldots = u_{in_i} = 0$ if $\gamma_i = 0$. 
From Theorem \ref{thm:perm_set} and by the definition of $\Gamma_i$s, it is not difficult to see that $u_{ij}\in [0,1]$ for all $ij$. 
Consider a random vector $A_{\bigcdot\bigcdot}$ that has the probability mass function in \eqref{eq:dist_A_set}, with $u_{ij}$s defined in \eqref{eq:u_ij_construct_set}. 
We can verity that 
$(A_{i1}, \ldots, A_{in_i})$s are mutually independent across all $i$, 
and the probability that $A_{ij}=1$ is proportional to 
\begin{align*}
	\exp(\gamma_i u_{ij}) =  \frac{\Pr(Y^0_{i1} = Y^0_{i(j)}\mid \mathcal{F}_{\text{set}})}{\min_{1\le k \le n_i}\Pr(Y^0_{i1} = Y^0_{i(k)}\mid \mathcal{F}_{\text{set}})}
	\propto \Pr(Y^0_{i1} = Y^0_{i(j)}\mid \mathcal{F}_{\text{set}}). 
\end{align*}
Consequently, 
$\sum_{k=1}^{n_i} A_{ik} q_{ik}$ equals $q_{ij}$ with probability $\Pr(Y^0_{i1} = Y^0_{i(j)}\mid \mathcal{F}_{\text{set}})$. 
From the proof of Theorem \ref{thm:perm_set}, for all $i\in [I]$,  
the distribution of 
$\sum_{k=1}^{n_i} A_{ik} q_{ik}$ is the same as the conditional distribution of $Q_{i1}$ given $\mathcal{F}_{\text{set}}$.
By the mutually independence, 
the distribution of 
$t(A_{\bigcdot\bigcdot}, Y^0_{\bigcdot(\bigcdot)}) = 
\sum_{i=1}^I \sum_{j=1}^{n_i} A_{ij} q_{ij}$ 
is the same as 
the conditional distribution of $t(Z_{\bigcdot \bigcdot}, Y_{\bigcdot \bigcdot}^0)=\sum_{i=1}^I Q_{i1}$ given $\mathcal{F}_{\textup{set}}$. 

From the above, Theorem \ref{thm:link_perm_rand_set} holds.

\subsection{Proof of Theorem \ref{thm:link_perm_rand_set_adap}}\label{sec:supp_adap_set}
From \eqref{eq:set_prob_ratio} and by the definition of $\tilde{\Gamma}_{\text{u}i}$s in \S \ref{sec:adapt_set}, for any $i\in [I]$ and $j,k\in [n_i]$, we have 
\begin{align}\label{eq:ratio_adap_supp}
	\frac{\Pr(Y_{i1}^0 = Y_{i(j)}^0\mid \mathcal{F}_{\textup{set}})/\hat{p}_{\text{m}ij}}{\Pr(Y_{i1}^0 = Y_{i(k)}^0 \mid \mathcal{F}_{\textup{set}})/\hat{p}_{\text{m}ik} }
	& 
	\in 
	\frac{ p_{\text{m}ij}/\hat{p}_{\text{m}ij}  }{
		p_{\text{m}ik}/ \hat{p}_{\text{m}ik}
	}
	\cdot 
	\big[ 
	\Gamma_{\text{u}i}^{-1}, \  
	\Gamma_{\text{u}i}
	\big]
	\subset 
	\big[ 
	\tilde{\Gamma}_{\text{u}i}^{-1}, \  
	\tilde{\Gamma}_{\text{u}i}
	\big]. 
\end{align}
For any $i\in [n_i]$, define 
\begin{align}\label{eq:u_ij_construct_set_adap}
	u_{ij} = 
	\frac{1}{\tilde{\gamma}_{\text{u}i}}
	\log \left\{ 
	\frac{
		\Pr(Y_{i1}^0 = Y_{i(j)}^0\mid \mathcal{F}_{\textup{set}})/\hat{p}_{\text{m}ij}
	}{
		\min_{1\le k \le n_i}\{ \Pr(Y_{i1}^0 = Y_{i(k)}^0\mid \mathcal{F}_{\textup{set}})/\hat{p}_{\text{m}ik} \}
	} \right\} 
	\qquad (j \in  [n_i])
\end{align}
if $\tilde{\gamma}_{\text{u}i} = \log \tilde{\Gamma}_{\text{u}i} > 0$, and $u_{i1} = \ldots = u_{in_i} = 0$ if $\tilde{\gamma}_{\text{u}i} = 0$. 
From \eqref{eq:ratio_adap_supp}, we can verify that $u_{ij}\in [0,1]$ for all $ij$. 
Consider a random vector $A_{\bigcdot\bigcdot}$ that has the probability mass function in \eqref{eq:cov_adap_set}, with $u_{ij}$s defined in \eqref{eq:u_ij_construct_set_adap}. 
We can verity that 
$(A_{i1}, \ldots, A_{in_i})$s are mutually independent across all $i$, 
and the probability that $A_{ij}=1$ is proportional to 
\begin{align*}
	\exp(\log \hat{p}_{\text{m}ij}+ \tilde{\gamma}_{\text{u}i} u_{ij}) & =
	\hat{p}_{\text{m}ij} \cdot \frac{
		\Pr(Y_{i1}^0 = Y_{i(j)}^0\mid \mathcal{F}_{\textup{set}})/\hat{p}_{\text{m}ij}
	}{
		\min_{1\le k \le n_i}\{ \Pr(Y_{i1}^0 = Y_{i(k)}^0\mid \mathcal{F}_{\textup{set}})/\hat{p}_{\text{m}ik} \}
	}
	\propto 
	\Pr(Y_{i1}^0 = Y_{i(j)}^0\mid \mathcal{F}_{\textup{set}}). 
\end{align*}
Following the same logic as the proof of Theorem \ref{thm:link_perm_rand_set}, we can know that 
the distribution of 
$t(A_{\bigcdot\bigcdot}, Y^0_{\bigcdot(\bigcdot)}) = 
\sum_{i=1}^I \sum_{j=1}^{n_i} A_{ij} q_{ij}$ 
is the same as 
the conditional distribution of $t(Z_{\bigcdot \bigcdot}, Y_{\bigcdot \bigcdot}^0)$ given $\mathcal{F}_{\textup{set}}$.

From the above, Theorem \ref{thm:link_perm_rand_set_adap} holds.

\subsection{Adaptive inference for matched set studies}\label{sec:adap_set_imple}

Below we give the details for the adaptive sensitivity analysis in \S \ref{sec:adapt_set}. 
Due to the equivalence between permutation and  randomization distributions in Theorem \ref{thm:link_perm_rand_set_adap}, 
the key step is to get the worst-case upper bound of the tail probability of the randomization distribution of 
$t(A_{\bigcdot\bigcdot}, Y^0_{\bigcdot(\bigcdot)}) = 
\sum_{i=1}^I \sum_{j=1}^{n_i} A_{ij} q_{ij}$ 
under the constraint that $\tilde{\Gamma}_{\text{u}i}\le \Lambda$ for all $i$ and some prespecified $\Lambda \ge 1$, 
where $A_{\bigcdot\bigcdot}$ follows the distribution in \eqref{eq:cov_adap_set} for some $u_{ij}$s in $[0,1]$. 
For each matched set $i$, 
let $q_{i(1)} \le \ldots \le q_{i(n_i)}$ be the sorted values of $(q_{i1}, \ldots, q_{i n_i})$, 
and, with a slight abuse of notation, 
let $\hat{p}_{\text{m}i(j)}$ be the corresponding value for each $q_{i(j)}$; 
that is, if $q_{i(j)} = q_{il}$ for some $l\in [n_i]$, then $\hat{p}_{\text{m}i(j)} = \hat{p}_{\text{m}il}$. 
Define further, for $i\in [I]$ and $1\le l \le n_i$, 
\begin{align*}
	\mu_{il} & = 
	\frac{
		\sum_{j=1}^l \hat{p}_{\text{m}i(j)} q_{i(j)} + \Lambda 
		\sum_{j=l+1}^{n_i} \hat{p}_{\text{m}i(j)} q_{i(j)}
	}{
		\sum_{j=1}^l \hat{p}_{\text{m}i(j)} + \Lambda 
		\sum_{j=l+1}^{n_i} \hat{p}_{\text{m}i(j)}
	}, 
	\\
	\nu_{il} & = 
	\frac{
		\sum_{j=1}^l \hat{p}_{\text{m}i(j)} q_{i(j)}^2 + \Lambda 
		\sum_{j=l+1}^{n_i} \hat{p}_{\text{m}i(j)} q_{i(j)}^2
	}{
		\sum_{j=1}^l \hat{p}_{\text{m}i(j)} + \Lambda 
		\sum_{j=l+1}^{n_i} \hat{p}_{\text{m}i(j)}
	}
	- \mu_{il}^2. 
\end{align*}
From \citet{pimentel2022covariate}, for each matched set $i$,
the maximum mean of $\sum_{j=1}^{n_i} A_{ij} q_{ij}$ and its maximum variance given that its mean is maximized have the following equivalent forms:
\begin{align*}
	\mu_{i} = \max_{l \in [n_i]} \mu_{il}, 
	\quad 
	\nu_{i} = \max_{l\in [n_i]: \mu_{il} = \mu_i } \nu_{il}.
\end{align*}
We can then construct an upper bound of the tail probability of $t(A_{\bigcdot\bigcdot}, Y^0_{\bigcdot(\bigcdot)}) = 
\sum_{i=1}^I \sum_{j=1}^{n_i} A_{ij} q_{ij}$ 
using the Gaussian distribution with mean $\sum_{i=1}^{I} \mu_{i}$ and variance $\sum_{i=1}^{I} \nu_{i}$; see \citet{GKR2000} and \citet{pimentel2022covariate} for more details. 

As a side note, the sensitivity analysis described in this subsection contains that in \S \ref{sec:sen_base_unif_perm} as a special case, where we can simply set all the $\hat{p}_{\text{m}ij}$s to be $1$.

\subsection{Extension to matched studies with one treated or one control unit per set}\label{sec:one_treat_or_control}

We consider more general matched observational studies where each matched set contains either (i)  one treated unit and one or multiple control units, or (ii) one control unit and one or multiple treated units, such as those from full matching. 
Below we show that the sensitivity analysis can be done similarly as that in \S \ref{sec:set}, by slightly modifying the biases $\Gamma_i$'s, the treatment assignments $Z_{ij}$'s and the $Q_{ij}$'s in the test statistic. This is analogous to Rosenbaum's approach 
to handling multiple treated units within a matched set; 
see, e.g., \citet[][pp. 161--162]{Rosenbaum02a}. 
Specifically, we consider a matched set $i$ with $n_i-1$ treated units and one control unit.
Without loss of generality, we assume that $Z_{i1} = 0$ and $Z_{i2} = \ldots = Z_{in_i} = 1$. The test statistic for this matched set can be equivalently written as 
$\sum_{j=1}^{n_i} Z_{ij} Q_{ij} = \sum_{j=1}^{n_i} \tilde{Z}_{ij} \tilde{Q}_{ij}$, 
with 
$\tilde{Z}_{ij} = 1 - Z_{ij}$ and 
$\tilde{Q}_{ij} = \sum_{k=1}^{n_i} Q_{ik} - Q_{ij}$. 
We can verify that, as long as the original $Q_{ij}$'s satisfy the property in Condition \ref{cond:symmetry_set}, the transformed $\tilde{Q}_{ij}$'s will also maintain this property. 
Therefore, we can essentially consider $\sum_{j=1}^{n_i} \tilde{Z}_{ij} \tilde{Q}_{ij}$, which contains only one ``treated'' unit after the deliberate label switching for this set, and apply the same methodology in \S \ref{sec:set}. 
Note that we need to be careful about the bias measure, since it generally changes after we switch the treatment labels. 
For example,  the bias measure $\Gamma_{\text{u}i}$ for a matched set with switched labels is defined analogously as in \eqref{eq:confound_strength_set} in \S \ref{sec:bias_set}, except that we need to switch the role of $\pi_1$ and $\pi_0$.

\end{document}